\documentclass{aa}  
\usepackage{graphicx}
\usepackage{txfonts}

\usepackage{natbib}
\usepackage[colorlinks=true, allcolors=blue]{hyperref}
\usepackage{multirow}

\def\sgras{Sgr~A$^\star$}
\def\chandra{{\it Chandra}}
\def\xmm{{\it XMM-Newton}}
\def\nustar{{\it NuSTAR}}
\def\swift{{\it Swift}-XRT}
\def\ep{{\it Einstein} Probe}

\def\xrism{XRISM}
\def\xtend{XRISM/Xtend}
\def\swiftxrt{{\it Swift}/XRT}

\usepackage{linenoaa}
\begin{document}

   \title{The very faint X-ray transient Swift J174610$-$290018 at the Galactic center}

   \subtitle{}

   \author{Giovanni Stel \inst{1,2}
          \and Gabriele Ponti \inst{1,3,2}
          \and Nathalie Degenaar\inst{4}
          \and Lara Sidoli\inst{5}
          \and Sandro Mereghetti\inst{5}
          \and Kaya Mori\inst{6}
          \and Tong Bao \inst{1}
          \and Giulia Illiano\inst{1} 
          \and Samaresh Mondal \inst{7}
          \and Mark Reynolds \inst{8,9}
          \and Chichuan Jin \inst{10,11,12}
          \and Tianying Lian \inst{10,11}
          \and Shifra Mandel\inst{6}
          \and Simone Scaringi\inst{13,14}
          \and Shuo Zhang \inst{15}
          \and Grace Sanger-Johnson \inst{15}
          \and Rudy Wijnands\inst{4}
          \and Jon M. Miller\inst{9}
          \and Jamie Kennea\inst{16}
          \and Zhenlin Zhu\inst{17}
          }

   \institute{
   INAF -- Osservatorio Astronomico di Brera, Via E. Bianchi 46, 23807 Merate, Italy \\
    \email{giovanni.stel@inaf.it}
    \and
Como Lake Center for Astrophysics (CLAP), DiSAT, Universit\`a degli Studi dell'Insubria, via Valleggio 11, I-22100 Como, Italy
    \and
    Max-Planck-Institut für extraterrestrische Physik, Giessenbachstrasse, 85748, Garching, Germany
    \and 
    Anton Pannekoek Institute for Astronomy, University of Amsterdam, NL-1098 XH Amsterdam, the Netherlands
    \and
    INAF - Istituto di Astrofisica Spaziale e Fisica Cosmica, Via A. Corti 12, I-20133 Milano,Italy
    \and
    Columbia Astrophysics Laboratory, Columbia University, New York, NY 10027, USA
    \and
    Department of Astronomy, University of Illinois, 1002 W. Green St., Urbana, IL 61801, USA
    \and
    Department of Astronomy, Ohio State University, 140 West 18th Ave., Columbus, OH 43210
    \and
    Department of Astronomy, The University of Michigan, 1085 South Univerity Avenue, Ann Arbor, MI, 48103, USA
    \and
    National Astronomical Observatories, Chinese Academy of Sciences, Beijing 100101, China
    \and
    School of Astronomy and Space Science, University of Chinese Academy of Sciences, Beijing 100049, China
    \and Institute for Frontier in Astronomy and Astrophysics, Beijing Normal University, Beijing 102206, China
    \and 
    Department of Physics, Centre for Extragalactic Astronomy, Durham University, South Road, Durham DH1 3LE, UK
    \and
    INAF -- Osservatorio Astronomico di Capodimonte, Salita Moiariello 16, I-80131 Naples, Italy
        \and
    Department of Physics and Astronomy, Michigan State University, East Lansing, 48824, United States
    \and
    Department of Astronomy and Astrophysics, Pennsylvania State University, 525 Davey Lab, University Park, PA 16802, USA
    \and 
    Department of Physics and Astronomy, University of California, Los Angeles, CA, 90095-1547, USA
             }

   \date{Received --- ; accepted ---}

 
  \abstract
   {Very faint X-ray transients (VFXTs) are a class of X-ray binary systems that exhibit occasional outbursts with peak X-ray luminosities ($L_X \lesssim 10^{36}$ erg s$^{-1}$) much lower than those of typical X-ray transients. On 22 February 2024, during its daily Galactic center monitoring, \swift\ detected a VFXT 7 arcmin from \sgras, dubbing it Swift J174610$-$290018.}
   {We aim to characterize the outburst that occurred in 2024 and a second, distinct outburst in 2025 to understand the nature and accretion flow properties of this new VFXT.}
   {\swift\ light curves were used to constrain the duration of the two events. We carried out X-ray spectral analysis, exploiting \xmm\ and \nustar\ data. We used \chandra\ and \xmm\ observations from the last 25 years to constrain the quiescent luminosity of the source and to compare the two most recent outbursts with previous detections of the source.}
   {During the 2024 outburst, which lasted about 50 days, the source reached a luminosity in the 2--10 keV band ($L_{2-10}$) of $\simeq 1.2 \times 10^{35}$ erg s$^{-1}$ (assuming it is located at the Galactic center, i.e., at a distance of 8.2 kpc). The 2025 outburst is shorter (about 5 days) and reached $L_{2-10}\simeq 9 \times 10^{34}$ erg s$^{-1}$. The spectral features of the source include an excess at 6.5--7 keV, which can be associated either with a single reflection line or with the ionized Fe XXV and XXVI lines. The same source was identified in both the \xmm\ and \chandra\ catalogs of point sources (known as 4XMM J174610.7$-$290020 and 2CXO J174610.7$-$290019). During previous detections the source displayed luminosity levels ranging from $L_{2-10}\simeq 2 \times 10^{32}$ to $L_{2-10}\simeq 3 \times 10^{33}$ erg s$^{-1}$ between 2000 and 2010. Moreover, it exhibited a potential type I X-ray burst in 2004.} 
   {The analysis of the outbursts and the potential type I burst strongly suggests that the VFXT Swift J174610$-$290018 is a
neutron star low-mass X-ray binary. The source can be described as being an accretion disk corona source (as has been recently proposed by an \xtend\ analysis). This scenario explains the overall low luminosity of this transient and the peculiar iron lines in the spectrum.}


   \maketitle

\section{Introduction}

The Galactic center (GC), being one of the most crowded and frequently observed regions of the Milky Way, has enabled the discovery of a few dozen X-ray transients \citep{Watson1981,Pavlinsky1994,Sidoli1999,Sidoli2001,Sakano2002,Muno2003,Muno2005,Muno2009,Wijnands2006,Kuulkers2007,Ponti2015, Degenaar2012,Degenaar2015, Mori2021}. Many of those sources are X-ray binaries that spend most of their time in quiescence with X-ray luminosities ($L_X$) of $\lesssim 10^{33}$~erg~s$^{-1}$ but occasionally show outbursts with peaks of $\sim 10^{37} - 10^{39}$~erg~s$^{-1}$ \citep{Degenaar2012}. A population of hundreds of X-ray binaries is believed to lie in the central parsec of the Milky Way \citep{Muno2005,Hailey2018, Mori2021}. Some of the X-ray transients exhibit peak luminosities considerably lower than those of typical X-ray binaries ($L_X \lesssim 10^{36}$ erg s$^{-1}$). These are usually referred to as very faint X-ray transients \citep[VFXTs;][]{Wijnands2006,Wijnands2015}.

A few dozen VFXTs have been detected in the Galaxy, and about a dozen in the GC \citep{Muno2005,Degenaar2012,Bahramian2023}.
The overabundance of these objects at the GC is probably due to an observation bias.
Indeed, the GC is the most frequently observed region of the Milky Way in the X-ray band. Almost every year, observations of the GC are performed with \chandra\ and \xmm\ (from 2000); \nustar\ (from 2012); and recently with the \ep\ (EP). Moreover, since 2006, the Swift X-Ray Telescope (\swift) has performed short ($\sim 1 $ ks) observations of the GC almost daily\footnote{Due to the Sun position, the GC is not observable between the end of November and the beginning of February.}, which allows the detection of transients down to $L_X \simeq 10^{34}$ erg s$^{-1}$ in the inner 15 arcmin of the Galaxy \citep{Degenaar2015}.

Given their typical low fluxes, the number of VFXTs that have been studied in detail is still modest; therefore, there is no clear, coherent picture yet of the demographics of this population. While the majority of VFXTs present outbursts that last for a few weeks \citep{Degenaar2015}, a few persistent and quasi-persistent VFXTs have been detected \citep{Bahramian2023}.
The VFXT class is heterogeneous, comprising sources of different natures. While a small fraction of VFXTs can be intrinsically bright, appearing faint due to inclination effects \citep{King2006}, most VFXTs are thought to be low-mass X-ray binaries (LMXBs) accreting at a low rate, thus explaining their relatively low luminosity \citep{Wijnands2006, Degenaar2015}.  Almost half of the studied VFXTs have been confirmed as LMXBs in which the compact object is a neutron star (NS) via the detection of type I X-ray bursts\footnote{Type I bursts are thermonuclear flashes on the surface of an accreting NS in a LMXB, caused by unstable burning of accumulated hydrogen and/or helium \citep{Lewin1993,Strohmayer2003}.} or coherent pulsations at the NS spin period; in a few cases, the binary system involves a black hole \citep[][]{Bahramian2023}. 

Various mechanisms have been introduced to explain the relatively low accretion rates of VFXTs. In the NS-LMXB case, a sufficiently strong magnetic field may quench the accretion flow via the magnetic propeller effect \citep{Degenaar2014, Degenaar2017,Heinke2015, Eijnden2018}. Alternatively, the low luminosity may be the result of a small accretion disk. This scenario would occur if the compact object accretes from a hydrogen-depleted or planetary companion in very compact orbits \citep{King2006,Hameury2016}, such as ultra-compact X-ray binaries. Studying new VFXTs
can reveal interesting subtypes of LMXBs, of which only a modest number have been discovered to date -- thereby tracing an accretion regime that is still relatively unexplored.

High-mass X-ray binaries (HMXBs) with X-ray luminosities ($L_X$) of $\lesssim 10^{36}$ erg s$^{-1}$,  can also belong to the VFXT class \citep{Wijnands2006,Mandel2025}. In this case, the compact object is typically accreting from the wind of the companion and an accretion disk might not be present. Moreover, accreting white dwarf cataclysmic variables (CVs) can contribute to the faint end of the VFXT population \citep{Hofmann2018}. In addition to binary systems, magnetars that have typical  quiescent luminosities ($L_X$) of $\sim10^{33}$ erg s$^{-1}$ \citep{Mereghetti2015} can also undergo faint outbursts consistent with VFXTs fluxes \citep{Coti_Zelati2018}; they make up a small fraction  of the VFXT population.

\begin{figure}
    \centering
    \includegraphics[width=\linewidth]{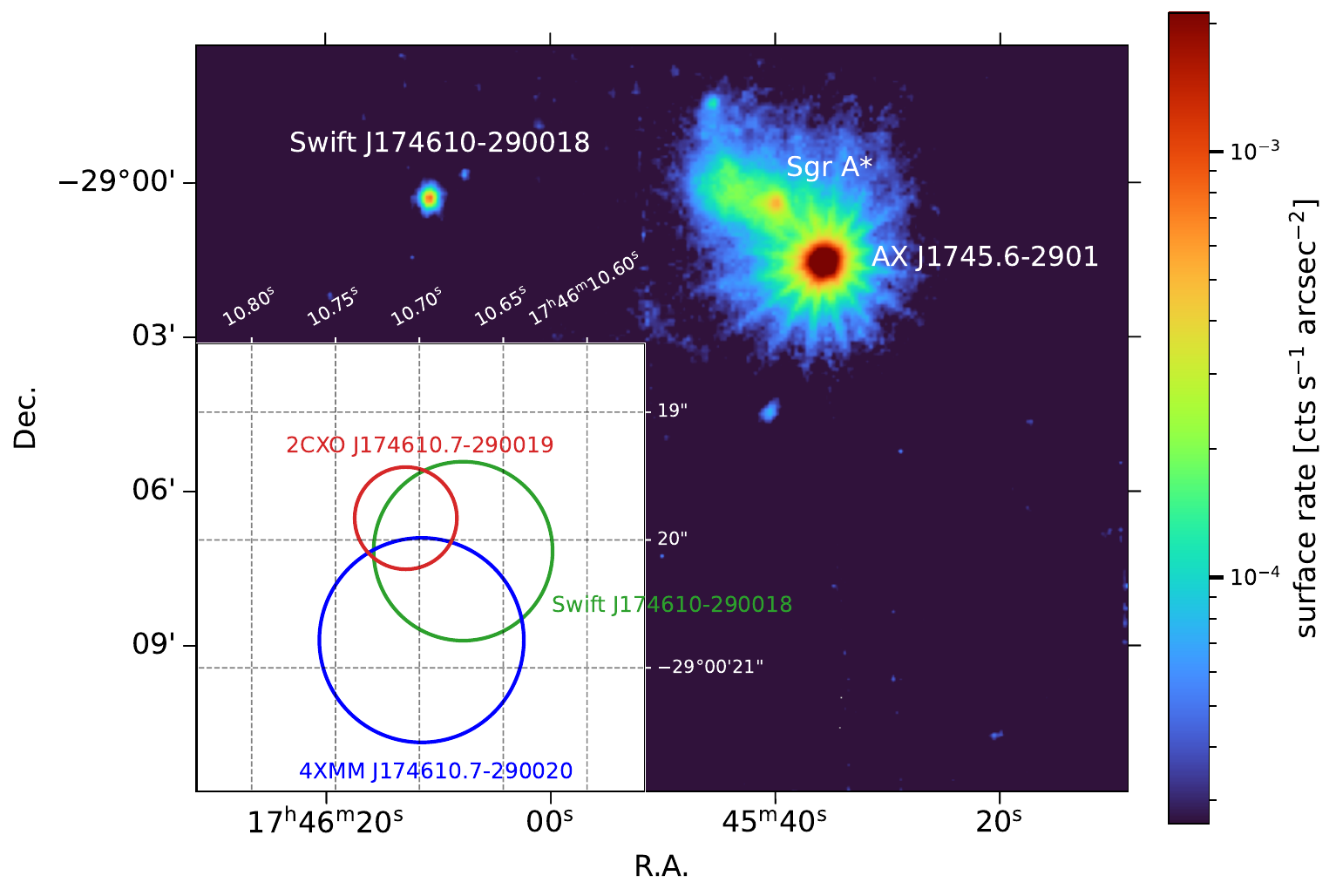}
    \caption{EPIC-pn image in the 2--10 keV band of the GC, obtained from the five \xmm\ observations taken in the spring of 2024. Swift J174610$-$290018 is  located 6.7 arcmin east of \sgras. The eclipsing NS-LMXB AX J1745--2901 \citep{Maeda1996} was active during the \xmm\ observations and is the brightest source in the field. The inset displays the position uncertainties ($1\sigma$) of the 2024 transient (green circle) and the known X-ray source reported in the 4XMM-DR14 and CSC 2.1 catalogs.}
    \label{fig:image}
\end{figure}

In this work we studied the outburst of an X-ray transient that was discovered by the \swift\ telescope on 22 February 2024 \citep{Reynolds2024} during the daily monitoring of the GC \citep{Degenaar2015}. The source, named Swift J174610$-$290018, was proposed as a new VFXT. The transient is located 6.7 arcmin east of \sgras\ (Fig.~\ref{fig:image}), corresponding to a projected distance of 16 pc at 8.2 kpc \citep{Gravity2019}, with an initial reported unabsorbed flux in the 2--10 keV band ($F_{2-10}$) of $(6.1 \pm 2.4) \times 10^{-12}$ erg cm$^{-2}$ s$^{-1}$, corresponding to a luminosity ($L_{2-10}$) of $\simeq 5 \times 10^{34}$ erg s$^{-1}$  at the GC distance \citep{Reynolds2024}. The outburst was also observed with \xmm, \nustar, and EP. \swift\ monitoring has shown that the source returned to quiescence at the beginning of April 2024 \citep{Degenaar2024}.

This transient was also observed on 26 February 2024 by the X-Ray Imaging and Spectroscopy Mission (\xrism) with the XTend instrument (the imaging camera); it showed a peculiar spectrum consistent with a NS-LMXB in addition to ionized iron lines (Fe XXV and Fe XXVI) at $\sim 7$ keV \citep{Yoshimoto2025}. The authors dubbed the source XRISM J174610.8$-$290021 and proposed it to be an accretion disk corona source \citep{White1982,Dove1997,Iaria2013} in which the direct emission from the NS-LMXB is blocked by the thick accretion flow, as the system is oriented edge-on. The observed flux is then the radiation-scattered by the ionized plasma that surrounds the accretion flow (the accretion disk corona).

\citet{Grefenstette2025} report a re-brightening of the same source, serendipitously observed with \nustar\  on 4 April 2025, with a spectrum and flux similar to those of the 2024 outburst. The source remained detectable in subsequent \nustar\ observations over the following few  days, before returning to quiescence \citep{Degenaar2025}.

The article is organized as follows: In Sect.~\ref{sec:dataset} we illustrate the dataset used in this study. In Sect.~\ref{sec:2024} we characterize the outburst that occurred in 2024. Section~\ref{sec:rebright} describes the outburst observed in 2025, and an archival search revealing previous detections of the source is addressed in Sect.~\ref{sec:past}. In Sect.~\ref{sec:discussion} we discuss the observational results.

\section{X-ray observations}
\label{sec:dataset}
A summary of the \xmm, \nustar, and \ep\ observations used for X-ray spectral analysis in this work is provided in Table~\ref{tab:list_obs}. The details for each telescope -- in addition to \swift\ and \chandra\ -- are presented below.

\begin{table*}[]
\caption{Log of the observations used for spectral analysis in this article.}
    \centering
    \begin{tabular}{c|c|c|c|c|c}
    \hline
       Telescope & Instrument (mode) &Obs. ID  &  Observation start  & Exposure &Notes\\
           &  &   &         [UTC]       & [s]      &       \\
    \hline
       \multirow{6}{*}{\xmm} & \multirow{6}{*}{EPIC-pn (Full Frame)}&0202670701 & 2004-08-31T03:37:41 & 135237 (69150)&type I burst?\\
        &&0944580601 & 2024-03-26T03:21:28 & 29800 (9500)&2024 outburst \\
        &&0944580701 & 2024-03-27T01:59:32 & 46200 (29300)&2024 outburst \\
        &&0944580801 & 2024-03-28T04:34:56 & 25800 (7200)&2024 outburst \\
        &&0944580501 & 2024-03-29T06:10:10 & 26400 (23000)&2024 outburst \\
        &&0932392301 & 2024-04-04T00:39:35 & 55682 (40300)&2024 outburst \\
        \hline
        \multirow{6}{*}{\nustar} &FPMA - FPMB& 30902013002 & 2024-04-04 05:13:22 & 20395& 2024 outburst\\
        &FPMA - FPMB& 31002004002 & 2025-04-04T10:35:06&17654 & 2025 outburst \\
        &FPMA - FPMB& 31002004004 & 2025-04-06T04:03:49& 18946& 2025 outburst \\
        &FPMA - FPMB& 31002004006 & 2025-04-07T03:59:28& 24843& 2025 outburst \\
        &FPMA - FPMB& 31002004008 & 2025-04-09T03:49:52& 18905& 2025 outburst \\
        \hline 
        \multirow{2}{*}{\ep} & FXT-A& \multirow{2}{*}{11911337728} &\multirow{2}{*}{2024-03-27T23:02:08}  & 31584& \multirow{2}{*}{2024 outburst}\\
        & FXT-B &  & & 31589& \\
        \hline
    \end{tabular}
     \tablefoot{For \xmm, the exposure reported in parentheses is the one after filtering for flaring particle background.}
    \label{tab:list_obs}
\end{table*}

\subsection{Swift}
We used \swift\ data of the GC monitoring campaign \citep{Degenaar2015} to create the light curves of the 2024 and 2025 outbursts. To this end, we used obsIDs 00096991167--00097302020 performed between 9 February and 21 April 2024, and obsIDs 00097302236--00097302245 performed between 2 and 11 April 2025. All observations were taken in photon counting mode and exposure times ranged from 0.70 to 1.21 ks. The raw \swift\ data were processed with the \textsc{xrtpipeline} using standard quality cuts and event grades 0–12. Light curves were created by extracting source events with \textsc{XSelect} using a 18-arcsecond circular region centered on the source position. 

\subsection{\xmm}
The outburst of the source is detected in five observations of the GC taken in the Spring of 2024 by \xmm\ \citep{Jansen2001}. We reprocessed the raw data using the standard pipelines in the Scientific Analysis System of \xmm\ (SAS 21.0), using the last release of the current calibration files as of April 2024. Given the higher effective area of EPIC-pn \citep{Struder2001} compared to the MOS \citep{Turner2001}, we used only the first camera in our analysis. We filtered the event list for high particle background by creating light curves of the entire CCD, in the 10--12 keV band, with a 100 s time bin. We then defined good time intervals as the periods in which the count rate is 
below 0.4 cts s$^{-1}$.

The spectra and light curves are extracted on a circular region of radius 12 arcsec. Background, unless otherwise specified, is accumulated in a nearby (RA, Dec= 17h46m03.43s, -29$^\circ$00'28.6'') circular region of 50 arcsec radius, where no point sources are evident. Spectra are collected from the filtered events list, with events selected using the SAS \texttt{evselect} task with the expression \texttt{"(FLAG==0) \&\& (PATTERN<=4)"}.
Light curves, instead, are computed from the event list, after the barycentric correction has been applied, not filtered for high particle background, to exploit the largest exposure available, using the SAS tool \texttt{evselect}. The expression in this case is \texttt{"\#XMMEA\_EP \&\& (PATTERN<=4)"}. Light curves are corrected for vignetting, and background subtracted with the task \texttt{epiclccorr}.

For the spectral analysis, best-fit parameters and related uncertainties are computed via Bayesian parameter estimation using Bayesian X-ray analysis \cite[BXA;][]{Buchner2014} for PyXspec \citep[based on Xspec version 12.13.0]{Arnaud1996}. Instead of subtracting the background, we modeled it, in the region indicated before, with a phenomenological model in Xspec: \texttt{tbabs*(apec+apec+gauss+gauss)+pow}. The two Gaussian lines account for the Fe K$\alpha$ (6.4 keV) line associated with molecular clouds in the region \citep{Stel2025}, and the Cu K$\alpha$ (8.0 keV) line is due to the internal "quiescent" background of the CCD\footnote{See \url{https://xmm-tools.cosmos.esa.int/external/xmm_user_support/documentation/uhb/epicintbkgd.html}}.

\subsubsection{Source position}
Given the source's count rate, its relative position on the CCD can be determined with a statistical uncertainty of approximately 0.1–0.2 arcsec. However, the overall positional uncertainty is dominated by the systematic error in \xmm\ absolute astrometry. To account for this, we retrieved the astrometrically corrected source coordinates from the Pipeline Processing System products for each observation, and performed a fit to determine the RA, Dec, and associated positional uncertainty, assuming Gaussian errors. The best position of the source is RA, Dec = (17h46m10.68s, -29$^{\circ}$00'20.1'') with a 0.7 arcsec uncertainty ($1 \sigma$). The position is fully consistent with the source 4XMM J174610.7$-$290020, as reported in the \textit{XMM-Newton} Serendipitous Source Catalogue \citep[4XMM-DR14;][]{Webb2020}. In Fig.~\ref{fig:image} we show the $1\sigma$ uncertainty of the source analyzed in this work (green circle) and the \xmm\ source reported in the catalog. 4XMM J174610.7$-$290020 was previously detected in four \xmm\ observations taken in 2000 and 2004. Given the consistent position, the fact that both sources are heavily absorbed and show similar spectral lines \citep{Pastor-Marazuela2020}, it is likely that they are the same source, observed in two different intensity states. For the remainder of this work, we consider them to be the same source.

\subsubsection{NIR counterpart}

We searched for a near-infrared (NIR) counterpart in the GALACTICNUCLEUS catalog \citep{Nogueras2021}.
Within 1.5 arcsec of the best source position reported before, there are 4 NIR sources (we excluded one foreground) from the GALACTICNUCLEUS catalog. Table~\ref{tab:nir_counterpart} reports the separation of the sources, their $K_s$ band magnitude, and, when available, the $H$ band magnitude.

\begin{table}[htb]
\caption{Potential NIR counterparts from the GALACTICNUCLEUS catalog.}
    \centering
    \begin{tabular}{c|c|c|c|c|c}
    \hline
         Ra & Dec & $\delta$ & $K_s$ & $H$ & $H-K$ \\
        $[$deg] & [deg] & [arcsec] & mag & mag & mag\\
         \hline
        266.54445 & -29.00553 & 0.20& 16.80 & --& --\\
        266.54459 & -29.00574 & 0.67& 15.43 & 17.84 & 2.41 \\
        266.54434 & -29.00578 & 0.86& 16.53 & --& --\\
        266.54417 & -29.00547 & 1.06&14.44 & 16.85& 2.41\\
        \hline
    \end{tabular}
     \tablefoot{$\delta$ is the angular separation between the NIR source and Swift J174610$-$290018. For all the NIR sources the uncertainty on RA and Dec is of the mas order, while the error on the magnitudes is typically 0.01 mag. }
    \label{tab:nir_counterpart}
\end{table}

Given the limitations of current NIR surveys, which are largely incomplete at $K_s\sim16$th mag due to crowding in the dense GC region, detecting an unevolved low-mass donor star is highly improbable.  We therefore consider it likely that none of the four NIR sources listed above is the true NIR counterpart to Swift J174610$-$290018.

\subsection{\chandra}

There are no \chandra\ \citep{Weisskopf2000} observations that cover the outburst in 2024. However, 
the same source is also reported in the \textit{Chandra} Source Catalog \citep[CSC 2.1,][]{Evans2024} with identifier 2CXO J174610.7$-$290019; as CXOU J174610.8$-$290019 in the \chandra\ catalog of the GC \citep{Muno2009} and in the Ultradeep \textit{Chandra} Catalog \citep[][source n. 3596]{Zhu2018}. 
According to the CSC 2.1, there are 11 observations between 1999 and 2024 in which the source has been detected with a significance > 3\footnote{The significance is the ratio of the photon flux measurement to its average error; see \url{cxc.cfa.harvard.edu/csc/columns/significance.html} for details.} in the ACIS-I 2.0--7.0 keV band. 

\subsection{\nustar}
\nustar\ \citep{Harrison2013} detected the source during the late stages of the 2024 outburst. However, given the relatively low flux, the \nustar\ analysis does not add a significant contribution to the \xmm\ observations of the outburst. On the other hand, the 2025 outburst was detected in four \nustar\ observations. We collected the spectra of the source in a circular region of radius 20 arcsec,  centered on the source. The background is accumulated in a 70 arcsec circular region at the coordinates: RA, Dec: (17h46m07.8s, -29$^\circ$01'53.3''). When analyzing the \nustar\ spectra, we fit the two modules FPMA and FPMB simultaneously, including a global multiplicative constant to account for cross-calibration uncertainties between the two instruments. We verified that the two spectra are overall consistent within a factor of 10\% in all observations.

\subsection{\ep\ (EP)}

The \ep\ detected the outburst of this source in late March 2024 during a performance verification observation of the GC, using the Follow-up X-ray Telescope (FXT) with an exposure time of $\sim$ 31.6 ks. EP-FXT comprises two co-aligned modules (FXT-A and FXT-B) in the 0.3-10 keV band, and the observation data of these two modules are processed using the FXT data analysis software (fxtsoftware v1.10). 

\section{2024 outburst}
\label{sec:2024}

\subsection{Light curves}
\begin{figure*}
    \includegraphics[width=\linewidth]{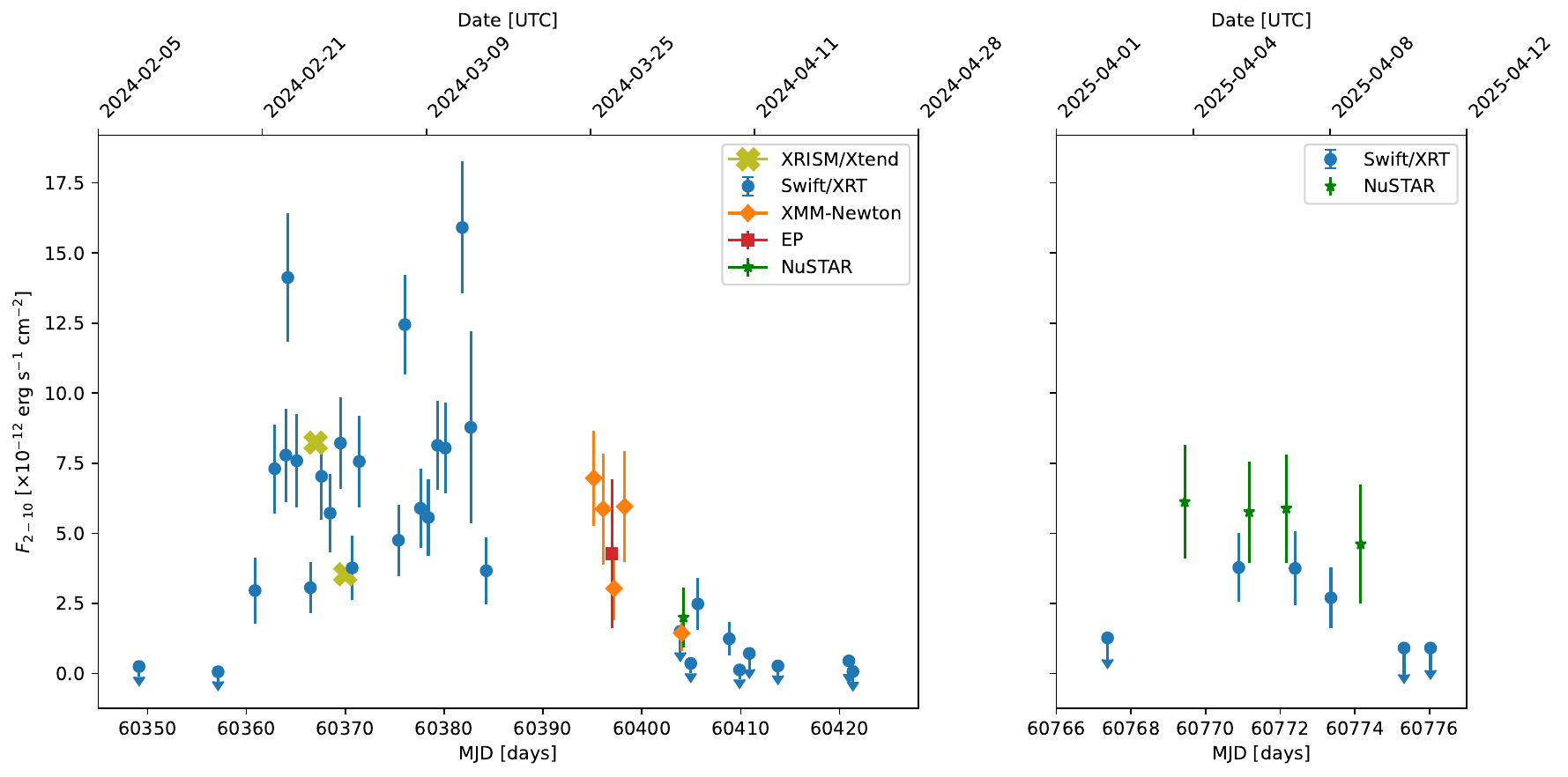}
    \caption{Light curve of the outburst in 2024 (left) and 2025 (right). $F_{2-10}$ is the 2--10 keV flux corrected for absorption, and it is computed assuming an absorbed power law spectrum, with $N_{\mathrm{H}} = 2 \times 10^{23}$ cm$^{-2}$ and $\Gamma=2$. The two \xrism\ data points were computed from the count rates reported in \citet{Yoshimoto2025}, assuming the same spectral model.}
    \label{fig:outburst}
\end{figure*}

\begin{figure*}
\centering
    \includegraphics[width=0.9\linewidth]{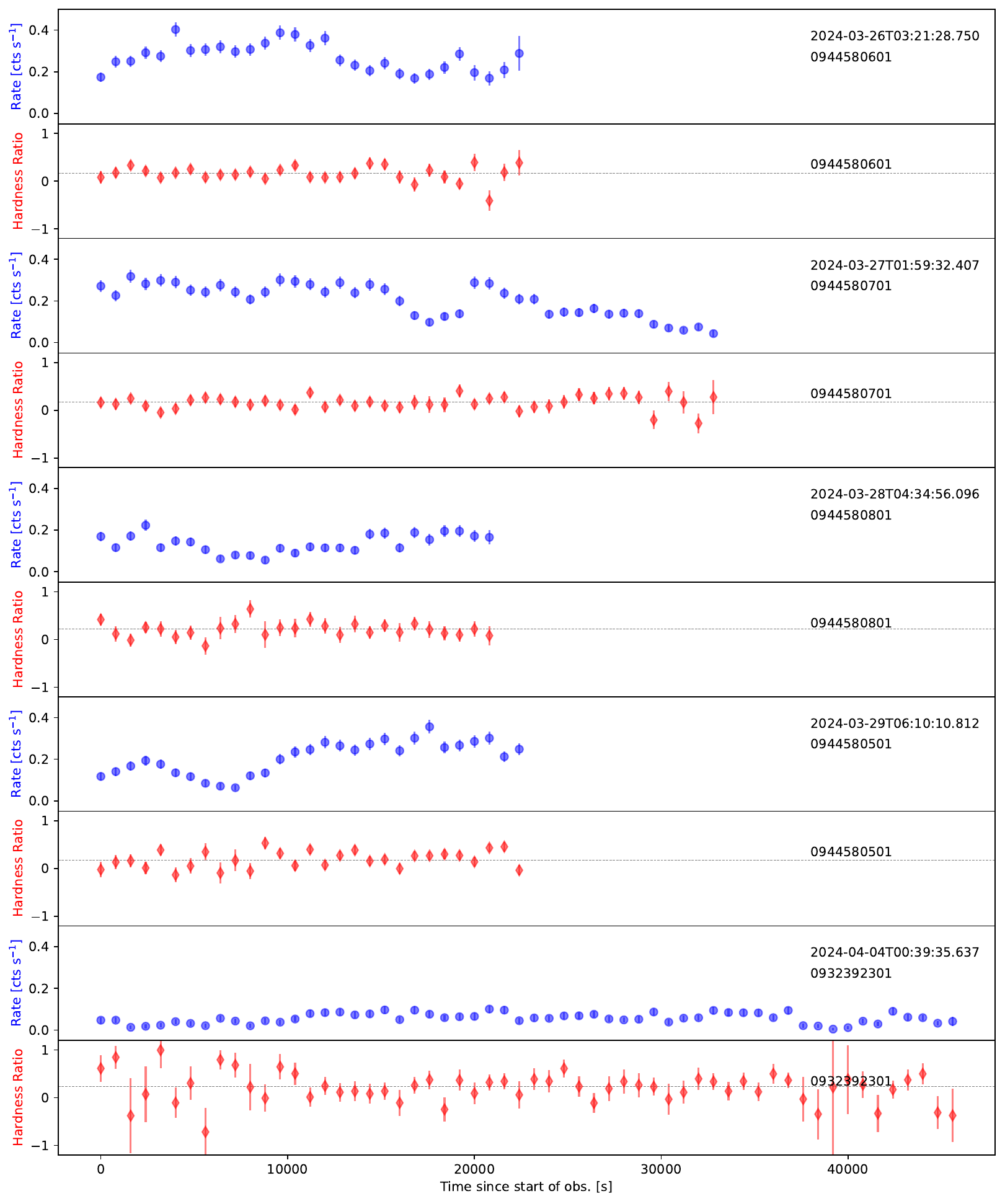}
    \caption{\xmm\ EPIC-pn light curves of the 2024 outburst (blue data) in the 2--10 keV band. Below each light curve, red data points are the hardness ratio computed from the light curves in the 2--5 and 5--10 keV energy bands. The dashed gray  line is the median hardness ratio for each observation.}
    \label{fig:lc}
\end{figure*}

The left panel of Fig.~\ref{fig:outburst} displays the light curve obtained during the outburst of the source in 2024. The count rate of the various instruments is converted to a 2--10 keV flux using \texttt{webpimms}\footnote{https://heasarc.gsfc.nasa.gov/cgi-bin/Tools/w3pimms/w3pimms.pl}, assuming a spectral model of an absorbed, power law with a photon index $\Gamma = 2$, and $N_{\mathrm H} = 2\times 10^{23}$ cm$^{-2}$ (see Sect.~\ref{sec:spectral_analysis}). The duration of the outburst is $\sim 50$ days. A gap in the \swift\ light curve is due to the fact that the satellite was in safe-mode from 15-03-2024 to 03-04-2024, for an issue related to the gyroscopes. 

\begin{table}[]
\caption{Timing analysis of the five \xmm\ observations.}
    \centering
    \begin{tabular}{c|c|c}
    \hline
       Date &  $\chi^2$ (n. data) & $\sigma$/mean  \\
         \hline
        2024-03-26  & 138 (29)& 0.26\\ 
        2024-03-27  & 548 (42)& 0.46\\
        2024-03-28  & 114 (27)& 0.37\\
        2024-03-29  & 354 (29)& 0.45\\
        2024-04-04  & 238 (58)& 0.55\\
        \hline
    \end{tabular}
    \tablefoot{The second column lists the $\chi^2$ relative to a fit using the weighted mean as a constant; in parenthesis is the number of data points.}
    \label{tab:td}
\end{table}

The 2--10 keV light curves of the five \xmm\ observations performed in spring 2024 are shown in Fig.~\ref{fig:lc} (blue), binned to 800 s. The last light curve, taken roughly six days after the others, exhibits a count rate reduced by a factor of about 4 compared to the previous four observations.  All five light curves show some variability. To test whether the variability is statistically significant, we calculated the $\chi^2$ with respect to the weighted mean of each observation. The results are shown in Table~\ref{tab:td}. In the same table, we report the ratio of the standard deviation and the weighted mean. Overall, the flux is fluctuating within 25-50\% of its mean value.
The hardness ratio\footnote{The hardness ratio (HR) is defined by $(H-S)/(H+S)$, where $H$ and $S$ are the count rates in the hard (5--10 keV) and soft (2--5 keV) bands, respectively.} between the 5--10 and 2--5 keV light curves shown in
Fig.~\ref{fig:lc} does not provide significant evidence of spectral variations.

\begin{figure}
    \centering
    \includegraphics[width=\linewidth]{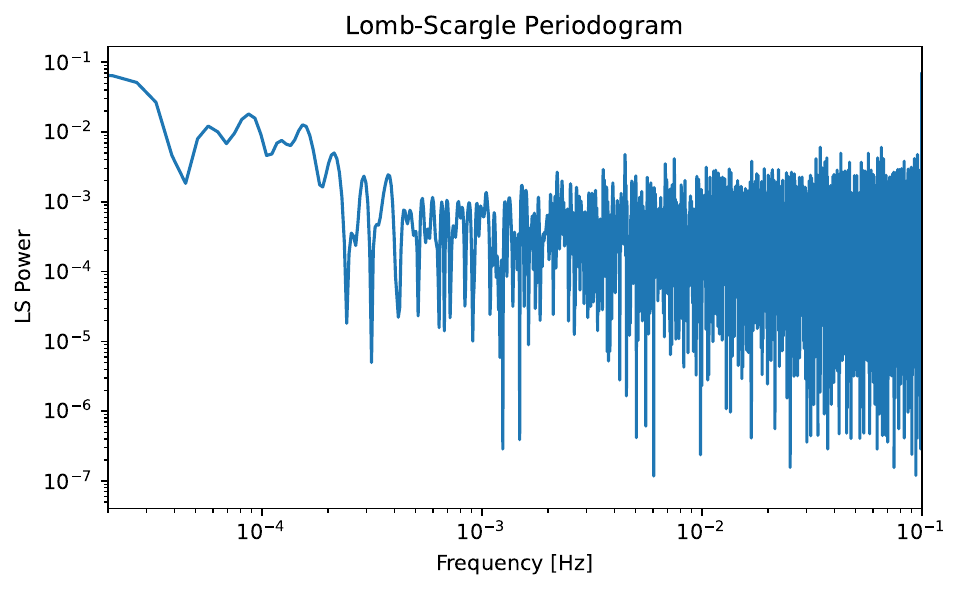}
    \caption{Lomb-Scargle periodogram of \xmm\ obs. 0944580701 (27 March 2024). The other observations show a similar behavior.}
    \label{fig:LS}
\end{figure}

We looked for periodicity of the emission in the 2--10 keV band by computing the Lomb-Scargle periodogram \citep{Lomb1976,Scargle1982} in the frequency range $4\times 10^{-5}-10^{-1}$ Hz. The periodogram of the longest \xmm\ observation  (ID. 0944580701, 2024-03-27) is shown as an example in Fig.~\ref{fig:LS}; the other periodograms have similar shapes. 
\citet{Yoshimoto2025} found a potential hint of periodicity of $\simeq 1537$ s, which we do not observe in the \xmm\ data.

\subsection{Spectral analysis}
\label{sec:spectral_analysis}

    \begin{figure*}
        \centering
        \includegraphics[width=\linewidth]{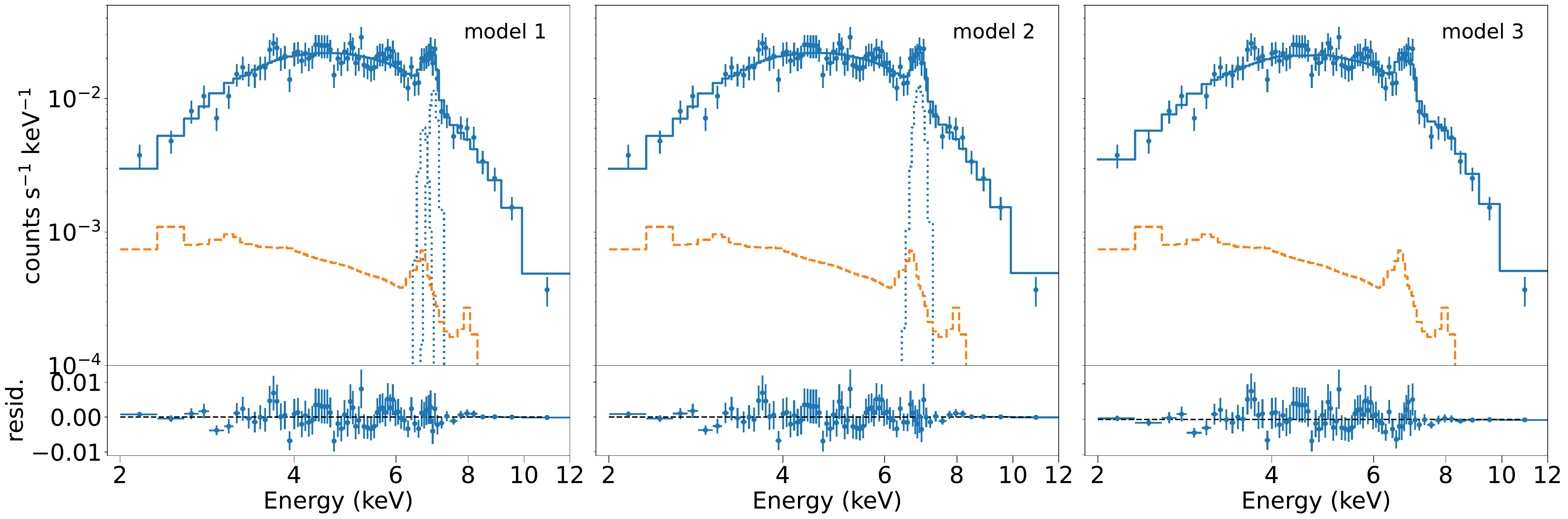}
        \caption{\xmm\ EPIC-pn spectrum of the source (blue) accumulated in Obs. 0944580701, with the three models discussed in Sect.~\ref{sec:spectral_analysis}. The spectra accumulated in the other four \xmm\ observations present similar shapes. Model 1 is an absorbed power law with the Fe XXV (6.7 keV) and Fe XXVI (6.97 keV) lines (dotted lines), while model 2 includes a single reflection line. Model 3 is ionized thermal plasma component (\texttt{apec}). The dashed orange line is the modeled background contribution.} 
        \label{fig:spectrum}
    \end{figure*}

Figure~\ref{fig:spectrum} displays the EPIC-pn spectrum of the longest \xmm\ observation (27 March 2024). The spectra of the other four \xmm\ observations, and the EP observations, present a similar shape. The spectrum is characterized by an excess at $\sim$ 7 keV. The most prominent emission lines in this region are the highly ionized iron lines Fe XXV ($\simeq 6.7$ keV) and Fe XXVI ($\simeq 6.97$ keV).

Since the five \xmm\ light curves exhibit different fluxes, and to maintain full generality, we fit each observation independently. As a first model ("model 1"), we implemented an X-ray-halo-scattered, absorbed power law with two Gaussian lines corresponding to the Fe XXV and Fe XXVI lines (\texttt{fgcdust*tbabs*(powerlaw+gaussian+gaussian)} in Xspec). The X-ray halo scattered model (\texttt{fgcdust}) is described in \citet{Jin2017,Jin2018}. This model is adopted since the source, as we will show, is heavily absorbed and in the projected direction of the GC; it is therefore likely a few kiloparsecs from the observer. This particular dust scattering model accounts for the dust material in the spiral arms of the Milky Way. We assumed abundances from \citet{Wilm2000}, and cross section from \citet{Verner1996}. 
The best parameters, along with their uncertainties, are reported in Table~\ref{tab:fit_pow}, while the spectrum is displayed in the left panel of Fig.~\ref{fig:spectrum}.
The source appears heavily absorbed with a column density typical of the sources located at the GC \citep[$N_{\mathrm H} \sim 10^{23}$ cm$^{-2}$,][]{Baganoff2003}. The photon index is consistent with $\Gamma \simeq 2$. The Fe XXVI line is the most prominent of the two iron lines and is consistently detected; while on two occasions we where able to only measure an upper limit for the flux of the Fe XXV line (see Table~\ref{tab:fit_pow}).

As an alternative model, since the Fe XXV and XXVI lines are not resolved (see Fig.~\ref{fig:spectrum}), we substituted the two ionized lines with a single line, allowing a non-null width ("model 2"; second panel in Fig.~\ref{fig:spectrum}). This is the case of a single reflection line from the accretion disk \citep{Fabian1989}. Table~\ref{tab:fit_pow} reports the fit result. Neither the $N_{\mathrm H}$ nor $\Gamma$ parameters were affected when switching from a model with two lines to that with a single one. The energy of the line is found to be $6.8-6.9$ keV, whereas the fit returns an upper limit on the width of $\simeq 200-300$ eV. The flux of the line is overall consistent among the first four observations, with a slight decrease in the last one.
In the same table, we report the difference of the Akaike information criterion (AIC) between model 2 and model 1. The model with the two ionized lines is slightly preferred, but the single reflection line cannot be ruled out.

Given the previous results, we attempted a fit of the spectrum with a collisionally ionized diffuse gas\footnote{A description of the \texttt{apec} model is available at \url{http://atomdb.org/}} (\texttt{fgcdust*tbabs*apec} in Xspec; "model 3"). Details of this fit are displayed in Table~\ref{tab:fit_apec}, and the spectrum is shown in the right panel of Fig.~\ref{fig:spectrum}. The measured absorption column density is lower
than the one inferred in the previous fits. 
The estimated temperature, which is set by the energy of the ionized iron lines and the continuum shape, is $kT\sim 12$ keV. In the table, we report the difference between the AIC of model 3 and model 1. 
Apart from the observation taken on 4 April 2024, there is no clear preference between the two models.

Overall, during the five \xmm\ observations, the source flux in the 2--10 keV band, corrected for absorption, is $F_{2-10} = (2-9) \times 10^{-12}$ erg s$^{-1}$ cm$^{-2}$, depending on the spectral model and of the observation date. 
\swiftxrt\ monitoring suggests that the source faded into the background level around the beginning of April 2024 (see Fig.~\ref{fig:outburst}).

\begin{table*}[]
\caption{Best-fit parameters and associated uncertainties ($1\, \sigma$) for the five \xmm\ spectra extracted during the 2024 outburst.}
    \centering
    \renewcommand{\arraystretch}{1.5}
    \begin{tabular}{c|c|c|c|c|c|c}
      & Parameter & 2024-03-26 & 2024-03-27 & 2024-03-28 & 2024-03-29 & 2024-04-04 \\
      \hline
      & $N_{\mathrm{H}}$ [$\times 10^{22}$ cm$^{-2}$] & $19 \pm 2$ & $18 \pm 2$ & $17 \pm 4$ & $19\pm 1$ & $14 \pm 2$ \\
      & $\Gamma$ & $2.1 \pm 0.2$ & $2.2 \pm 0.2$ & $2.2 \pm 0.2$ & $1.9 ^{+0.2}_{-0.1}$ & $1.7 \pm 0.2$ \\
      & log norm. [ph/keV/cm$^2$/s] & $-2.4\pm 0.2$ & $-2.6 \pm 0.1$ & $-2.7 \pm 0.3$ & $-2.7\pm0.1$ & $-3.5\pm 0.2$ \\
      \hline
      \multirow{2}{*}{model 1} & log $F_{6.7}$ [ph/cm$^2$/s] & $-5.1^{+0.2}_{-0.4}$ & $-5.3^{+0.2}_{-0.3}$ & $<-4.9^*$ & $<-5.3^*$ & $-5.4 \pm 0.1$ \\
       & log $F_{6.97}$ flux [ph/cm$^2$/s] & $-4.9 \pm 0.2$ & $-4.9 \pm 0.2$ & $-5.0^{+0.2}_{-0.3}$ & $-5.0 \pm 0..1$ & $-5.5 \pm 0.2$ \\
      \hline
      \multirow{4}{*}{model 2} & E [keV] & $6.84 \pm 0.09$& $6.89 \pm 0.03$ & $6.82^{+0.06}_{-0.08}$ & $6.95 \pm 0.06$ & $6.82^{+0.06}_{-0.05}$\\
      & $\sigma$ [eV] & $<300^*$ & $<130^*$ & $<200^*$ & $<220^*$ & $<250^*$\\
      & log $F_{\mathrm{line}}$ [ph/cm$^2$/s] & $-4.7 \pm 0.2$& $-4.79_{-0.09}^{+0.08}$ & $-4.7\pm 0.2$ & $-4.9^{+0.1}_{-0.2}$ & $-5.1 \pm 0.1$\\
      & $\Delta$ AIC$_{2-1}$ & 4.1& 1.7 & 1.3  & 4.4 & 3.8\\ 
      \hline
      & $F_{2-10}$ [$\times 10^{-12}$ erg s$^{-1}$ cm$^{-2}$] & $8.8^{+1.1}_{-0.9}$ & $5.6^{+0.5}_{-0.4}$ & $4.6^{+0.9}_{-0.7}$ & $6.4_{-0.4}^{+0.5}$ & $1.6^{+0.2}_{-0.1}$ \\
    \end{tabular}
    \tablefoot{Model 1 includes the Fe XXV and Fe XXVI lines, while model 2 a single reflection line. $F_{2-10}$ is the de-absorbed flux in the 2-10 keV band, assuming model 1.\\
    $^*$: upper limit. We report the 84$^{\text{th}}$ percentile of the posterior.}
    \label{tab:fit_pow}
\end{table*}

\begin{table*}[]
\caption{Same as Table~\ref{tab:fit_pow} but with an \texttt{apec} model instead of a power law model.}
    \centering
    \renewcommand{\arraystretch}{1.5}
    \begin{tabular}{c|c|c|c|c|c|c|c}
        & Parameter & 2024-03-26 & 2024-03-27 & 2024-03-28 & 2024-03-29 & 2024-04-04 \\
        \hline
        \multirow{5}{*}{model 3} & $N_{\mathrm{H}}$ [$\times 10^{22}$ cm$^{-2}$] & $14.7 \pm 1.0$ & $14.1^{+0.7}_{-0.6}$ & $14 \pm 2$ & $16.6 \pm 0.8$ & $14.5 \pm 1.4$ \\
        & $kT$ apec [keV] & $12^{+3}_{-2}$ & $12^{+2}_{-1}$ & $10^{+4}_{-2}$ & $16 \pm 3$ & $10^{+2}_{-1}$ \\
        &norm. [$\times 10^{-3}$ cm$^{-5}$] & $5.3 \pm 0.3$ & $3.5 \pm 0.2$ & $3.0 \pm 0.3$ & $4.1 \pm 0.2$ & $1.20_{-0.08}^{+0.09}$ \\
        &$F_{2-10}$ [$\times 10^{-12}$ erg s$^{-1}$ cm$^{-2}$] & $7.2_{-0.4}^{+0.3}$  &$4.7^{+0.1}_{-0.2}$ & $3.9 \pm 0.3$ & $5.6 \pm 0.2$ & $1.59 \pm 0.08$ \\
        &$\Delta$ AIC$_{3-1}$ & -2.5 & 2.1 & -2.7 & -1.6 & -5.2 \\
    \end{tabular}
    \label{tab:fit_apec}
\end{table*}

\section{Re-brightening of 2025}
\label{sec:rebright}

On 4 April 2025, \nustar\ serendipitously detected a re-brightening of the source \citep{Grefenstette2025}. The new outburst was confirmed the following day by \swift. The light curve of this second outburst is displayed in the right panel of Fig.~\ref{fig:outburst}. The source is clearly detected between 4 and 9 April 2025 in four \nustar\ and three \swift\ observations. The source was not detected by \swift\ on 2 April 2025, and in the days after 10 April 2025. Therefore, the duration of this second outburst is significantly shorter than the 2024 event, elapsing only $\simeq 5$ days. 

\begin{figure*}
        \centering
        \includegraphics[width=0.8\linewidth]{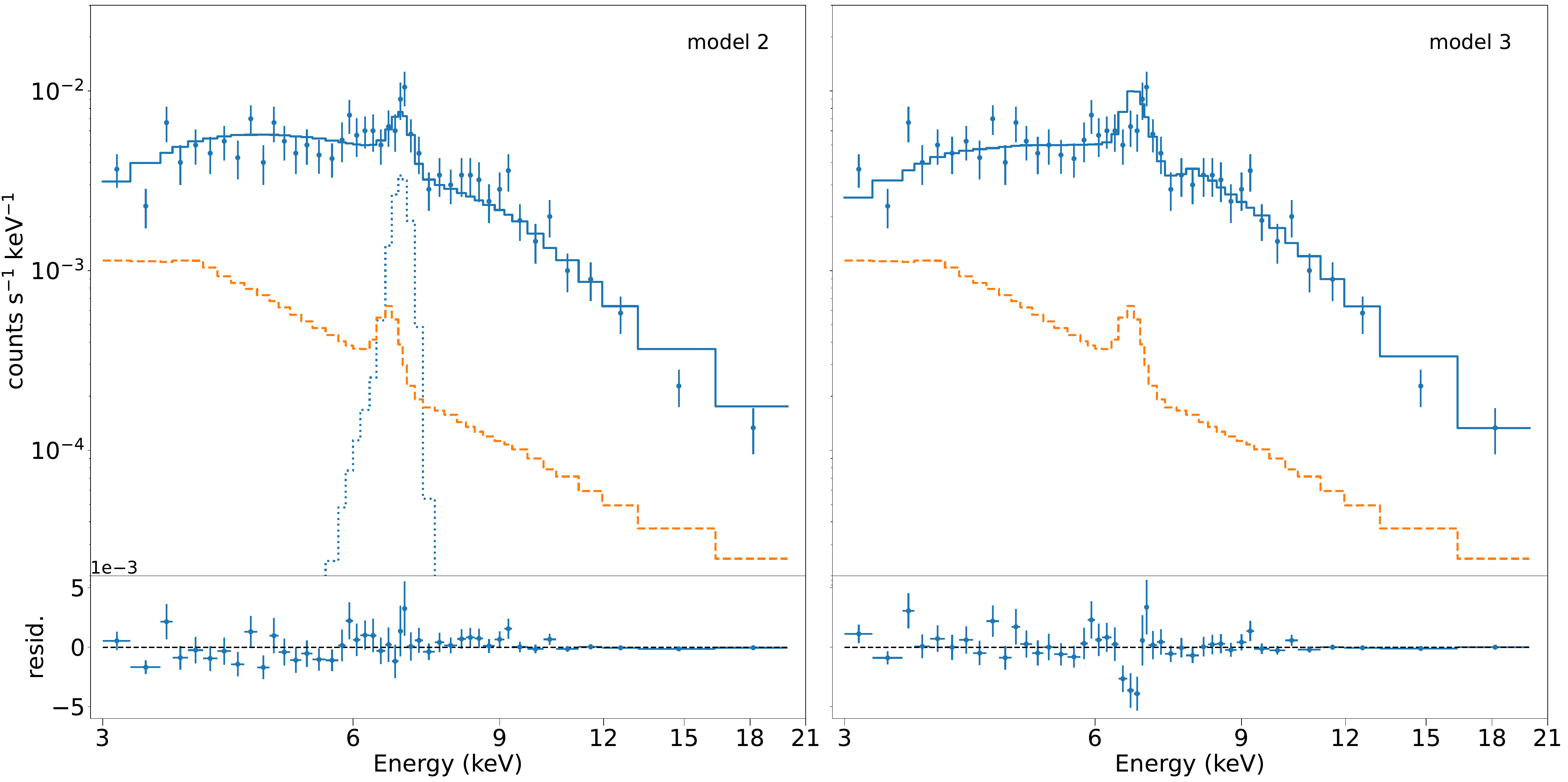}
        \caption{\nustar\ FPMB spectrum of the source (blue) accumulated in Obs. 31002004006, with the two models discussed in Sect.~\ref{sec:rebright}. Model 2 includes a single reflection line (dotted line) with a power law continuum. Model 3 is an ionized thermal plasma component (\texttt{apec}). The dashed orange line is the background contribution.} 
        \label{fig:spectrum_2025}
    \end{figure*}

We collected the spectrum during the four \nustar\ observations in which the transient was detected, and performed the fit in the 3-20 keV band. Given that the \nustar\ spectral resolution is lower than \xmm, we fit only the model 2 and model 3, without attempting to distinguish between a single or a combination of two iron lines. Figure~\ref{fig:spectrum_2025} shows the \nustar\ FPMB spectrum taken on 7 April 2025 (ID. 31002004006), along with the two models. The best-fit parameters are reported in Table~\ref{tab:nustar2025}. To allow a direct comparison between the parameters of the 2024 outburst, and since \nustar\ is not sensitive below 3 keV, we fixed the column density to the best values found in the previous outburst. Overall, the spectrum appears softer when compared with the 2024 outburst (Table~\ref{tab:fit_pow} and \ref{tab:fit_apec}). If fitted with a power law, the photon index is close to $\Gamma \simeq 3$, while fitted with an \texttt{apec} component, the temperature is $kT\simeq 5-6$ keV. The normalization of the Fe line is consistent between the two outbursts. The AIC difference clearly shows a preference for model 2 (power law + line). The \texttt{apec} model, on the other hand, struggles to reproduce simultaneously the iron lines and the continuum.

\begin{table*}[]
    \caption{Best spectral fit parameters of the 2025 outburst, observed with \nustar.}
    \centering
    \renewcommand{\arraystretch}{1.5}
    \begin{tabular}{c|c|c|c|c|c}
    \hline
        & Parameter & 2025-04-04 & 2025-04-06 & 2025-04-07 & 2025-04-09 \\
        \hline
        \multirow{8}{*}{model 2} & $N_{\mathrm{H}}$ [$\times 10^{22}$ cm$^{-2}$] & $18^\dag$ & $18^\dag$ & $18^\dag$ & $18^\dag$ \\
        & $\Gamma$ & $3.14 \pm 0.09$ & $3.05 \pm 0.08$& $2.89^{+0.07}_{-0.08}$& $2.89^{+0.10}_{-0.09}$\\
        & log norm. [ph/keV/cm$^2$/s] & $-1.48^{+0.08}_{-0.07}$ & $-1.55 \pm 0.07$ & $-1.72^{+0.06}_{-0.07}$ & $-1.79^{+0.09}_{-0.08}$ \\
        & E [keV] & $6.84 \pm 0.09$& $6.67^{+0.13}_{-0.11}$ & $6.87 \pm 0.05$ & $6.74 \pm 0.12$ \\
        & $\sigma$ [eV] & $50^\dag$ & $50^\dag$ & $50^\dag$ & $50^\dag$\\
        & log $F_{\mathrm{line}}$ [ph/cm$^2$/s] & $-4.8^{+0.2}_{-0.4}$& $-4.8_{-0.6}^{+0.2}$ & $-4.54\pm^{+0.09}_{-0.10}$ & $-4.9^{+0.2}_{-0.6}$\\
        & const & $1.08\pm0.07$& $1.14 \pm 0.07$& $1.08 \pm 0.06$&$1.08 \pm 0.08$ \\
        & $F_{2-10}$ [$\times 10^{-11}$ erg s$^{-1}$ cm$^{-2}$] & $1.8\pm 0.1$ & $1.7\pm 0.1$ & $1.5\pm 0.1$ & $1.2 \pm 0.1$ \\
        \hline
        \multirow{6}{*}{model 3} & $N_{\mathrm{H}}$ [$\times 10^{22}$ cm$^{-2}$] & $15^\dag$ & $15^\dag$ & $15^\dag$ & $15^\dag$ \\
        & $kT$ apec [keV] & $5.1^{+0.4}_{-0.3}$ & $5.3^{+0.4}_{-0.3}$ & $6.1 \pm 0.4$ & $6.3^{+0.6}_{-0.5}$  \\
        &norm. [$\times 10^{-3}$ cm$^{-5}$] & $10.7^{+1}_{-0.9}$ & $10.0^{+0.9}_{-0.7}$ & $8.5 \pm 0.6$ & $6.8 \pm 0.6$  \\
        & const & $1.08^{+0.07}_{-0.06}$& $1.14^{+0.08}_{-0.07}$& $1.08 \pm 0.06$&$1.09 \pm 0.08$ \\
        &$F_{2-10}$ [$\times 10^{-11}$ erg s$^{-1}$ cm$^{-2}$] & $1.08 \pm 0.06$  &$1.08 \pm 0.06$ & $0.95 \pm 0.05$ & $0.76 \pm 0.05$  \\
        & $\Delta$ AIC$_{3-2}$ & -33 &-32 &-11 & -18\\ 
        \hline
    \end{tabular}
    \tablefoot{ $^\dag$ fixed.
    }
    \label{tab:nustar2025}
\end{table*}

\section{Past outbursts and detections}
\label{sec:past}

\begin{figure}
    \centering
    \includegraphics[width=\linewidth]{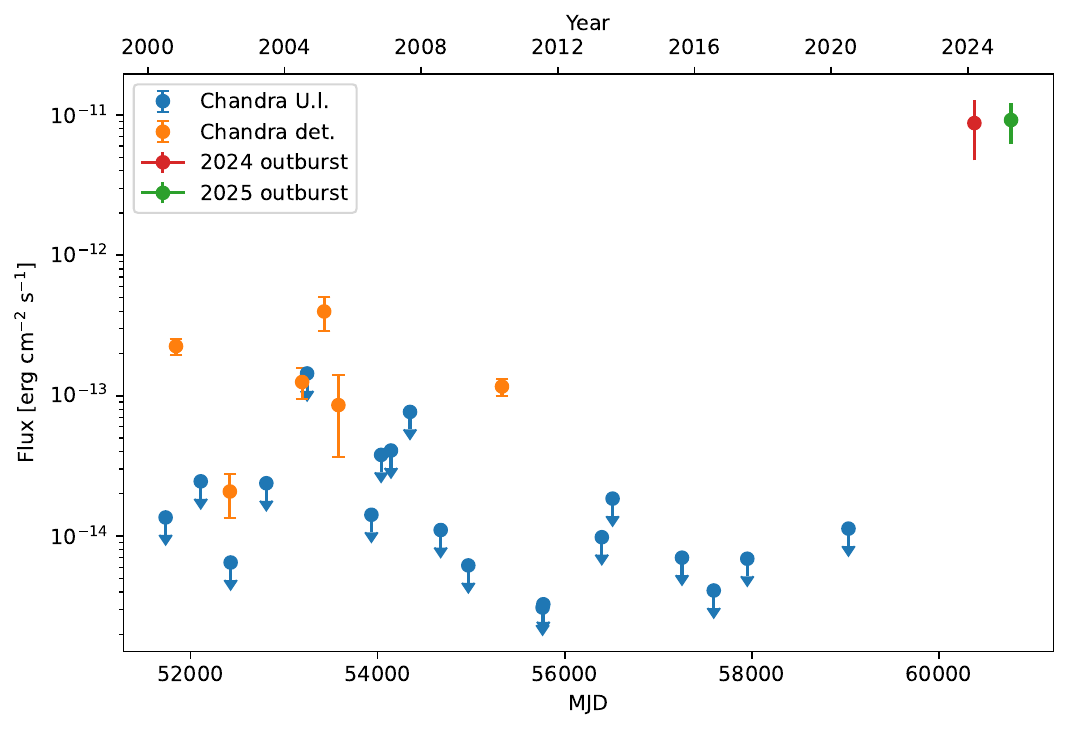}
    \caption{Long-term light curve of the source studied in this work.} 
    \label{fig:long_lc}
\end{figure}

In Fig.~\ref{fig:long_lc} we present the light curve of the source from 2000 onward, compiled using archived observations from \chandra. Upper limits (blue data points) and detections (orange) are retrieved from the CSC 2.1. The conversion from count rate to a flux is performed with the same model used in Fig.~\ref{fig:outburst} ($N_{\mathrm{H}} = 2 \times 10^{23}$ cm$^{-2}$ and $\Gamma=2$). For graphical purposes, consecutive observations that are within 120 days are merged. The mean flux is shown in cases of detection, while only the more stringent upper limit is shown in cases of non-detection. The source has been detected by \chandra\ a few times, with a 2--10 keV flux within $2\times 10^{-14}$ erg s$^{-1}$ cm$^{-2}$ and $4\times 10^{-13}$ erg s$^{-1}$ cm$^{-2}$. The reported upper limits indicate a quiescence flux $F_{2-10} \le 3 \times 10^{-15}$ erg s$^{-1}$ cm$^{-2}$.

\subsection{Potential type I burst}
\begin{figure}
    \centering
    \includegraphics[width=\linewidth]{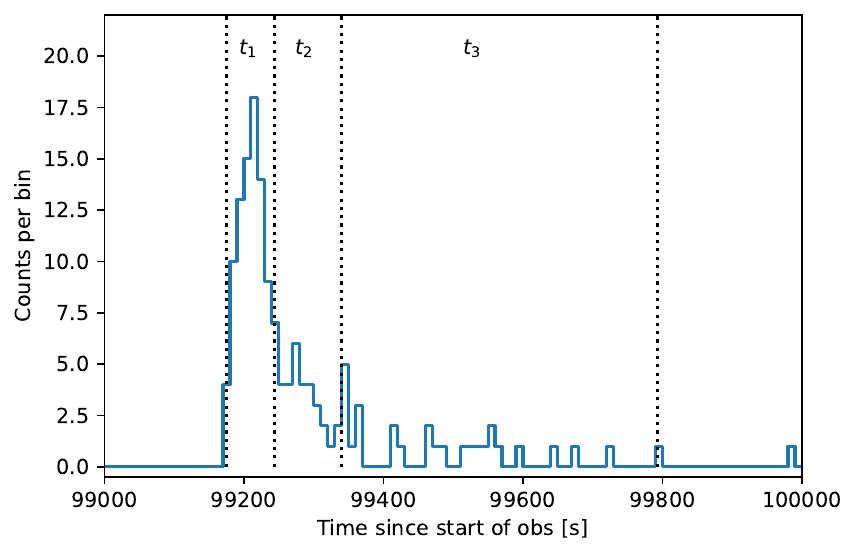}
    \caption{\xmm\ EPIC-pn (full frame mode) light curve of the 2004 flare in the 2--10 keV band. The temporal bin-size is 10 s. The dotted vertical lines separate the three intervals used for the time-resolved spectral analysis.}
    \label{fig:flare_100}
\end{figure}

During an \xmm\ observation performed on 31 August 2004  (ID n. 0202670701), in which the source was detected along the whole observations, the light curve of the source displayed a potential type I burst \citep{Pastor-Marazuela2020}. Figure~\ref{fig:flare_100} displays the histogram of the events (bin-size of 10 s) during the burst.
To determine the duration of the burst, we performed a Bayesian-block decomposition \citep{Scargle2013} on the EPIC pn unfiltered event list associated with the source. We used the \texttt{bayesian\_blocks} function from the Python package \texttt{Astropy}, assuming $p0=0.01$, where $p0$ gives the false alarm probability, which is used to compute the prior. The burst is clearly identified for a total duration of 523 s. The duration is longer than typical X-ray type I bursts \citep[from a few seconds to about a minute,][]{Galloway2008}, but consistent with ``intermediate-duration bursts" that last a few to tens of minutes \citep{Alizai2023}. The peak of the burst is reached after $\simeq 40$ s. 

\begin{figure}
    \centering
    \includegraphics[width=\linewidth]{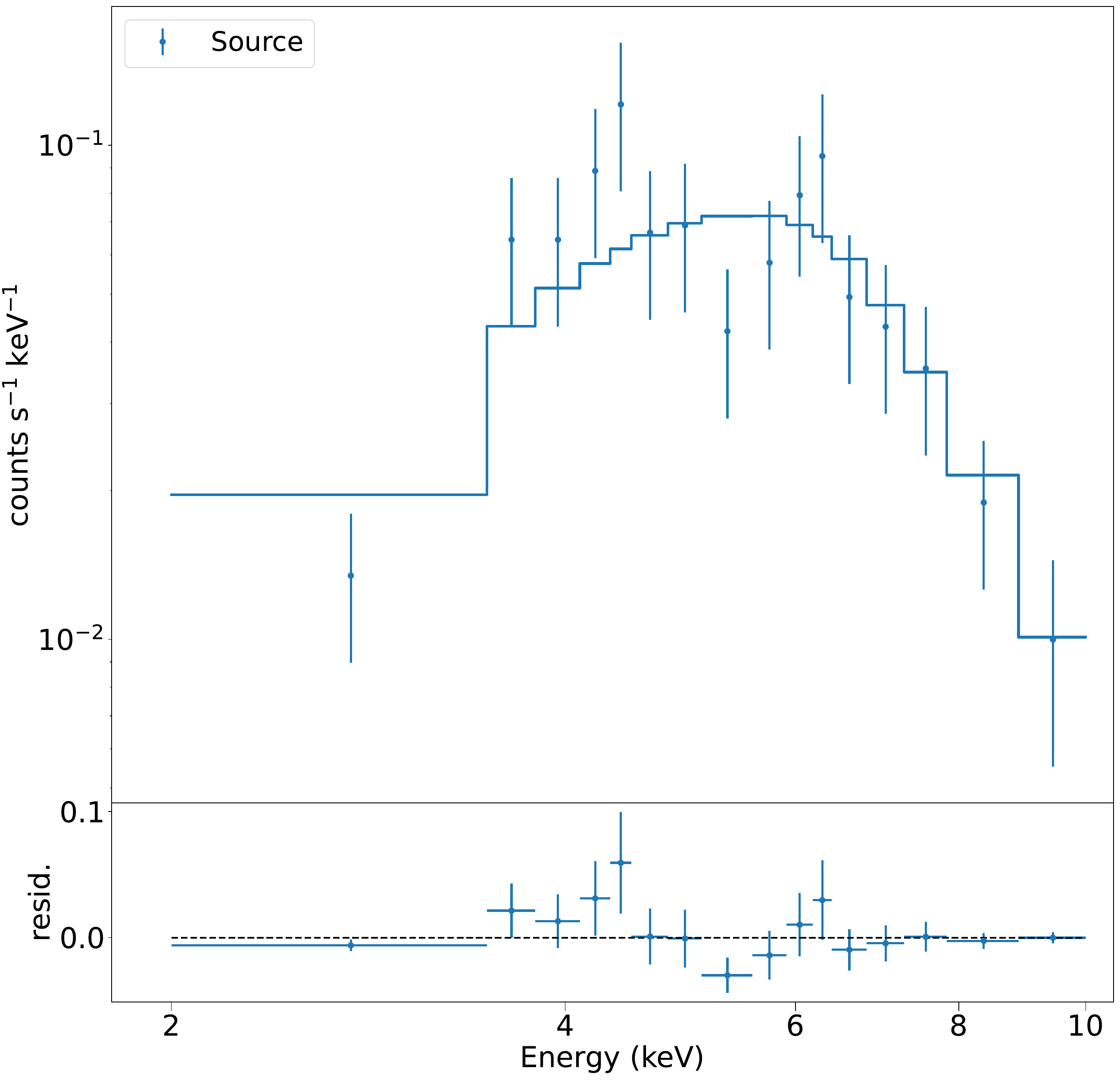}
    \caption{\xmm\ EPIC-pn spectrum of the potential type I burst that occurred in 2004. The model is an absorbed black body (see Table~\ref{tab:typeI}).}
    \label{fig:fit_flare}
\end{figure}

We integrated the spectrum during the flare interval in the same region as before.  We estimated the background by accumulating a spectrum in the same position, before the start of the burst feature. We chose a phenomenological model for the background (instrumental and pre-burst emission) consisting of an absorbed thermal plasma component, with a power law and a 6.4 keV line (\texttt{tbabs*(apec+pow+gauss)} in Xspec).
The model for the burst is an absorbed black body emission (\texttt{fgcdust*tbabs*bbodyrad} in Xspec). The spectrum, with the associated fit, is shown in Fig.~\ref{fig:fit_flare}. The derived temperature is $kT = 2.8^{+1.4}_{-0.7}$ keV, for an absorption column density of $N_{\mathrm H} = (7 \pm 5) \times 10^{22}$ cm$^{-2}$. 
The mean de-absorbed flux in the 2-10 keV band is $F_{2-10} = 1.9^{+0.3}_{-0.2}\times 10^{-11}$ erg s$^{-1}$ cm$^{-2}$. For a given distance ($d$), the normalization of the black body spectrum can be converted to the source radius ($R$) as\begin{equation}
R=0.2^{+0.1}_{-0.07} \times \frac{d}{8.2 \, \text{kpc}} \quad \text{[km].} 
\end{equation}

To search for any spectral softening during the burst, we divided it into three intervals highlighted in Fig.~\ref{fig:flare_100}. We then fit the spectrum in each interval; the best parameters are reported in Table~\ref{tab:typeI}. A softening of the spectrum is identified, consistent with the typical behavior of  type I  X-ray bursts. To verify that the spectrum is not excessively affected by the pile-up effect during the $t_1$ interval, we performed the analysis again, selecting only single-pattern events ("PATTERN==0" in the SAS \texttt{evselect} function). The results are consistent within the statistical uncertainties.

Type I bursts that reach Eddington luminosity can be exploited as standard candles to determine the distance to the NS-LMXB. This kind of type I burst presents photospheric radius expansion and has a bolometric peak luminosity ($L_{\mathrm{bol}}$) of $\simeq 3.8 \times 10^{38}$ erg s$^{-1}$ \citep{Kuulkers2003}. The burst of 2004, for which there is no evidence of photospheric radius expansion, is much fainter. The peak lies within the $t_1$ interval, as shown in Fig.~\ref{fig:flare_100}, and the count rate of the peak is about twice the mean count rate measured during $t_1$. Assuming that the source is located at the GC (distance of 8.2 kpc), the luminosity in the 2--10 keV band during $t_1$ is $L_{2-10}= (7 \pm 1)\times10^{35} $ erg s$^{-1}$, which translates into a bolometric peak luminosity of $L_{\mathrm{bol}}\sim 10^{37} $ erg s$^{-1}$, computed in the energy range 0.1--100 keV.

\begin{table*}[]
\caption{Best-fit parameters for the potential type I burst.}
    \centering
    \renewcommand{\arraystretch}{1.5}
    \begin{tabular}{c|c|c|c|c}
    \hline
      Parameter  & Total & $t_1$ & $t_2$ & $t_3$\\
      \hline
      Duration [s]& 523 & 69& 95& 359 \\
      
      $N_{\mathrm{H}}$ [$\times 10^{22}$ cm$^{-2}$]  & $7 \pm 5$ &$7^\dag$ & $7^\dag$&$7^\dag$\\
      
      $kT$ [keV]  & $2.8^{+1.4}_{-0.7}$ & $4.0^{\flat}$& $2.6^{+1.5}_{-0.7}$&$1.2^{+0.3}_{-0.2}$\\
      
      log norm [keV]  & $-1.2 \pm 0.4$& $-1.2^{+0.3}_{-0.2}$& $-1.0^{+0.4}_{-0.5}$& $-0.7 \pm 0.4$\\
      
      $F_{2-10}$ [$\times 10^{-11}$ erg s$^{-1}$ cm$^{-2}$] & $1.9^{+0.3}_{-0.2}$& $9\pm 1$& $2.4_{-0.4}^{+0.6}$&$0.40 \pm 0.08$\\
      
      \hline
    \end{tabular}
    \tablefoot{$^\dag$: fixed.\\
    $^\flat$: lower limit. We report the 16$^{\text{th}}$ percentile of the posterior.}
    \label{tab:typeI}
\end{table*}

\section{Discussion}
\label{sec:discussion}
Swift J174610$-$290018 underwent two distinct outbursts in the last two years. Assuming that the source is located at the GC, the fluxes reported in Sects.~\ref{sec:2024} and \ref{sec:rebright} translates into a luminosity of $L_{2-10} \simeq(1-12)\times 10^{34}$ erg s$^{-1}$ and $L_{2-10} \simeq(6-9)\times 10^{34}$ erg s$^{-1}$ for the 2024 and 2025 outburst, respectively. These two outbursts are the brightest ever recorded for this source; therefore, the transient belongs to the VFXT class. The same source was detected in the past at lower fluxes by \chandra\ and \xmm. The flux in the previous detections is in the range $F_{2-10}\simeq (0.2-4)\times 10^{-13}$ erg cm$^{-2}$ s$^{-1}$ (see Fig.~\ref{fig:long_lc}), corresponding to a luminosity $L_{2-10}\simeq (0.2-3)\times 10^{33}$ erg s$^{-1}$, assuming the source is located at the GC. Upper limits instead indicates a  quiescence luminosity of $L_{2-10}\le 2\times 10^{31}$ erg s$^{-1}$.
Our spectral analysis of the 2024 outburst is in agreement with the results of \citet{Yoshimoto2025}. When fitted with the same model (apec or power law continuum), the parameters are consistent (i.e., $kT \sim 12$ keV and $\Gamma \sim 2$), showing that there is no sign of spectral evolution during the outburst.  

Establishing the nature of this source is not straightforward. The spectrum is characterized by a spectral slope of $\Gamma \sim 2-3$, which is common for VFXT that are NS-LMXBs \citep{Degenaar2009,Wijnands2015}. Our analysis does not answer whether the iron complex emission is due to a single reflection line or the sum of ionized Fe XXV and Fe XXVI lines. In any case, iron lines are unusual for VFXTs, which are NS-LMXBs, with only a few cases being observed \citep{Eijnden2018,Delsanto2007}. For example, \citet{Eijnden2018} reported for the VFXT IGR J17062--6143 the presence of a reflection iron line in the spectrum, which could arise in the case of an ultra-compact X-ray binary or a magnetically truncated disk. 

The study of the potential type I burst, is a strong indication of the NS-LMXB nature of the source. Indeed, the flare observed in 2004 shows several key features typical of type I bursts: the light curve (a fast rise followed by an exponential decay), the spectral shape (a blackbody with a temperature of a few keV), and the spectral softening during the burst \citep{Galloway2008,Degenaar2018}. Nevertheless, the duration of a few hundred seconds is longer than typical type I bursts, and falls in the intermediate-duration bursts class. Those bursts are related to NS-LMXB accreting below the Eddington limit (i.e., $< 0.03 \, \dot M_{\mathrm{Edd}}$), and present  a recurrence time of months to years rather than the hours to days of regular bursts \citep{Alizai2023}.

A coherent explanation of these features can be achieved assuming that this particular VFXT is an accretion disk corona source, as proposed by \citet{Yoshimoto2025}. In this scenario the NS-LMXB is seen almost edge on, and the radiation from the source is blocked by the thick accretion disk. The only observed flux is the emission scattered by the accretion disk corona, while the iron lines are due to the ionized plasma. In this case, both the outburst and the type I burst are intrinsically brighter, but only the fainter, scattered emission is observed. In principle, if the NS–LMXB system is viewed almost edge-on, as proposed in this scenario, eclipses should be periodically visible in the light curve. The lack of detected eclipses may be due to the limited exposure time of the observations. In addition, the accretion disk corona can be larger than the line of sight obscured by the donor passage. For instance, for a NS-LMXB, with NS of  $ 1.4 \, M_\odot$, and donor star of mass $0.5 \, M_\odot$, the Roche lobes are respectively $R_{L1}\simeq 0.8 \, R_\odot$ and $R_{L2}\simeq 0.5 \, R_\odot$ \citep{Eggleton1983}. Assuming that the extension of the accretion disk corona is roughly the circularization radius ($\approx 0.8\, R_{L1}$), the donor star radius is less than it. Therefore, the passage of the donor star never entirely obscure the extended source. Alternatively, since the type I burst implies that the source is a NS-LMXB, the lack of eclipses or dips would imply that the system is not observed edge-on and, therefore, that the source and the type I burst are intrinsically faint. The iron complex excess can then be a reflection disk line.

Other physical scenarios can explain part of the characteristics of this source, but struggle to reproduce all of them coherently.
For example, HMXBs are another class of sources that exhibit ionized iron lines due to wind-fed accretion. HMXBs with X-ray luminosities comparable to those observed during the outburst of the last two years are known \citep{Lutovinov2013,Mandel2025}. However, it remains challenging to explain the burst feature with a softening tail that occurred during the \xmm\ observations of 2004.

CVs also present a thermal spectrum with strong ionized iron lines. In particular, the observed X-ray luminosity can be reached by an intermediate polar (IP). IPs are magnetic CVs that have unsynchronized orbits. The white dwarf's magnetic field strength is $B \sim (0.1-10)\times 10^6 \,\text{G}$; between polars ($B \sim (10-300)\times 10^6 \, \text{G}$), and nonmagnetic CVs  ($B \lesssim  10^5 \, \text{G}$). However, IPs are usually persistent and more luminous \citep[$L_{2-10} \sim \times 10^{33}$ erg s$^{-1}$;][]{Mukai2017,Xu2019}; therefore, they cannot explain the observed broad variability of about 3 orders of magnitude and a quiescent luminosity of  $< 2 \times 10^{31}$ erg s$^{-1}$. In principle, a white dwarf can experience such luminosity variations in the case of a nova\footnote{A nova is a thermonuclear runaway of H-rich material on the surface of a white dwarf.} \citep{Mukai2017}. This scenario has been recently proposed by \citet{Hua2025}. However, even if the two events observed in 2024 and 2025 were indeed novae, the rapid flare (type I burst like) observed in 2004 remains difficult to explain, strongly suggesting that the source is a NS-LMXB.

Regarding the search for a NIR counterpart, it is inconclusive, as low mass main-sequence stars are practically non detectable because of source confusion at faint magnitudes. Regarding the four NIR sources that lie within the positional uncertainty of Swift J174610$-$290018, using the extinction value of each source reported in \citet{Nogueras2021} and assuming a distance of 8.2 kpc, the absolute K-magnitude of the four sources is in the range between $M_K\simeq -2.7$  and $M_K\simeq -0.5$. These magnitudes are typical of giants (see \citealt{Nogueras2021} for details), or type-B stars \citep{Mandel2025}. The X-ray analysis suggests that the source is a NS-LMXB; therefore, the NIR counterpart is likely undetected.

If the low luminosity observed during the 2024 and 2025 events results from the accretion disk corona scenario, then Swift J174610$-$290018 would be classified as a VFXT due to its inclination rather than to a low accretion rate. However, as Fig.~\ref{fig:long_lc} shows, previous detections at fluxes 1–3 orders of magnitude lower suggest that the source also experienced phases of reduced accretion. For example, during the 2004 type I burst, immediately preceding the flare, the source was nearly two orders of magnitude fainter than during the 2024 and 2025 outbursts. The low accretion state in that period may therefore explain the unusually long duration of that type I burst \citep{Alizai2023}.

\section{Conclusions}

We analyzed the two distinct outbursts of the GC VFXT Swift J174610$-$290018 (also known as XRISM J174610.8$-$290021) that occurred in 2024 and 2025. These represent the brightest outbursts ever observed from this source; they lasted approximately 50 and 5 days, respectively, with a typical X-ray luminosity in the 2--10 keV band ($L_{2-10}$) of $\simeq(1-12)\times 10^{34}$ erg s$^{-1}$. In both events, the source exhibited spectra characterized by a strong iron complex. An accretion disk corona, as proposed by \citet{Yoshimoto2025}, can account for the transient's low luminosity and its spectral shape. We studied a short flare from the same source that occurred in 2004, which shows the characteristics of type I X-ray bursts, thereby confirming its NS-LMXB nature. Although the accretion disk corona scenario can account for the relatively low luminosity of the recent outbursts, the source has previously been observed at luminosities up to three orders of magnitude lower, indicating intrinsically fainter outbursts and a broad variability in its accretion behavior.

\begin{acknowledgements}
We thank the anonymous referee for their comments that
helped improve the quality of the paper. This work is based on observations obtained with \textit{XMM-Newton}, an ESA science mission with instruments and contributions directly funded by ESA Member States and NASA. This project acknowledges financial support from the European Research Council (ERC) under the European Union’s Horizon 2020 research and innovation program HotMilk (grant agreement No. 865637), support from Bando per il Finanziamento della Ricerca Fondamentale 2022 dell’Istituto Nazionale di Astrofisica (INAF): GO Large program and from the Framework per l’Attrazione e il Rafforzamento delle Eccellenze (FARE) per la ricerca in Italia (R20L5S39T9). S.Mereghetti acknowledges financial support through the INAF grants ``Magnetars'' and ``Toward Neutron Stars Unification''. SM acknowledges support from the grant NASA ADAP 80NSSC24K0666. CJ acknowledges the National Natural Science Foundation of China through grant 12473016, and the support by the Strategic Priority Research Program of the Chinese Academy of Sciences (Grant No. XDB0550200). LS acknowledges funding from the grant entitled ``Bando Ricerca Fondamentale INAF 2023"  
\end{acknowledgements}

\bibliographystyle{aa}
\bibliography{sample}

@ARTICLE{Harrison2013,
       author = {{Harrison}, Fiona A. and {Craig}, William W. and {Christensen}, Finn E. and {Hailey}, Charles J. and {Zhang}, William W. and {Boggs}, Steven E. and {Stern}, Daniel and {Cook}, W. Rick and {Forster}, Karl and {Giommi}, Paolo and {Grefenstette}, Brian W. and {Kim}, Yunjin and {Kitaguchi}, Takao and {Koglin}, Jason E. and {Madsen}, Kristin K. and {Mao}, Peter H. and {Miyasaka}, Hiromasa and {Mori}, Kaya and {Perri}, Matteo and {Pivovaroff}, Michael J. and {Puccetti}, Simonetta and {Rana}, Vikram R. and {Westergaard}, Niels J. and {Willis}, Jason and {Zoglauer}, Andreas and {An}, Hongjun and {Bachetti}, Matteo and {Barri{\`e}re}, Nicolas M. and {Bellm}, Eric C. and {Bhalerao}, Varun and {Brejnholt}, Nicolai F. and {Fuerst}, Felix and {Liebe}, Carl C. and {Markwardt}, Craig B. and {Nynka}, Melania and {Vogel}, Julia K. and {Walton}, Dominic J. and {Wik}, Daniel R. and {Alexander}, David M. and {Cominsky}, Lynn R. and {Hornschemeier}, Ann E. and {Hornstrup}, Allan and {Kaspi}, Victoria M. and {Madejski}, Greg M. and {Matt}, Giorgio and {Molendi}, Silvano and {Smith}, David M. and {Tomsick}, John A. and {Ajello}, Marco and {Ballantyne}, David R. and {Balokovi{\'c}}, Mislav and {Barret}, Didier and {Bauer}, Franz E. and {Blandford}, Roger D. and {Brandt}, W. Niel and {Brenneman}, Laura W. and {Chiang}, James and {Chakrabarty}, Deepto and {Chenevez}, Jerome and {Comastri}, Andrea and {Dufour}, Francois and {Elvis}, Martin and {Fabian}, Andrew C. and {Farrah}, Duncan and {Fryer}, Chris L. and {Gotthelf}, Eric V. and {Grindlay}, Jonathan E. and {Helfand}, David J. and {Krivonos}, Roman and {Meier}, David L. and {Miller}, Jon M. and {Natalucci}, Lorenzo and {Ogle}, Patrick and {Ofek}, Eran O. and {Ptak}, Andrew and {Reynolds}, Stephen P. and {Rigby}, Jane R. and {Tagliaferri}, Gianpiero and {Thorsett}, Stephen E. and {Treister}, Ezequiel and {Urry}, C. Megan},
        title = "{The Nuclear Spectroscopic Telescope Array (NuSTAR) High-energy X-Ray Mission}",
      journal = {\apj},
         year = 2013,
        month = jun,
       volume = {770},
       number = {2},
          eid = {103},
        pages = {103},
          doi = {10.1088/0004-637X/770/2/103},
archivePrefix = {arXiv},
       eprint = {1301.7307},
 primaryClass = {astro-ph.IM},
       adsurl = {https://ui.adsabs.harvard.edu/abs/2013ApJ...770..103H}
}

@INPROCEEDINGS{Weisskopf2000,
       author = {{Weisskopf}, Martin C. and {Tananbaum}, Harvey D. and {Van Speybroeck}, Leon P. and {O'Dell}, Stephen L.},
        title = "{Chandra X-ray Observatory (CXO): overview}",
    booktitle = {X-Ray Optics, Instruments, and Missions III},
         year = 2000,
       editor = {{Truemper}, Joachim E. and {Aschenbach}, Bernd},
       series = {Society of Photo-Optical Instrumentation Engineers (SPIE) Conference Series},
       volume = {4012},
        month = jul,
        pages = {2-16},
          doi = {10.1117/12.391545},
archivePrefix = {arXiv},
       eprint = {astro-ph/0004127},
 primaryClass = {astro-ph},
       adsurl = {https://ui.adsabs.harvard.edu/abs/2000SPIE.4012....2W}
}

@ARTICLE{Turner2001,
       author = {{Turner}, M.~J.~L. and {Abbey}, A. and {Arnaud}, M. and {Balasini}, M. and {Barbera}, M. and {Belsole}, E. and {Bennie}, P.~J. and {Bernard}, J.~P. and {Bignami}, G.~F. and {Boer}, M. and {Briel}, U. and {Butler}, I. and {Cara}, C. and {Chabaud}, C. and {Cole}, R. and {Collura}, A. and {Conte}, M. and {Cros}, A. and {Denby}, M. and {Dhez}, P. and {Di Coco}, G. and {Dowson}, J. and {Ferrando}, P. and {Ghizzardi}, S. and {Gianotti}, F. and {Goodall}, C.~V. and {Gretton}, L. and {Griffiths}, R.~G. and {Hainaut}, O. and {Hochedez}, J.~F. and {Holland}, A.~D. and {Jourdain}, E. and {Kendziorra}, E. and {Lagostina}, A. and {Laine}, R. and {La Palombara}, N. and {Lortholary}, M. and {Lumb}, D. and {Marty}, P. and {Molendi}, S. and {Pigot}, C. and {Poindron}, E. and {Pounds}, K.~A. and {Reeves}, J.~N. and {Reppin}, C. and {Rothenflug}, R. and {Salvetat}, P. and {Sauvageot}, J.~L. and {Schmitt}, D. and {Sembay}, S. and {Short}, A.~D.~T. and {Spragg}, J. and {Stephen}, J. and {Str{\"u}der}, L. and {Tiengo}, A. and {Trifoglio}, M. and {Tr{\"u}mper}, J. and {Vercellone}, S. and {Vigroux}, L. and {Villa}, G. and {Ward}, M.~J. and {Whitehead}, S. and {Zonca}, E.},
        title = "{The European Photon Imaging Camera on XMM-Newton: The MOS cameras}",
      journal = {\aap},
         year = 2001,
        month = jan,
       volume = {365},
        pages = {L27-L35},
          doi = {10.1051/0004-6361:20000087},
archivePrefix = {arXiv},
       eprint = {astro-ph/0011498},
 primaryClass = {astro-ph},
       adsurl = {https://ui.adsabs.harvard.edu/abs/2001A&A...365L..27T}
}

@ARTICLE{Struder2001,
       author = {{Str{\"u}der}, L. and {Briel}, U. and {Dennerl}, K. and {Hartmann}, R. and {Kendziorra}, E. and {Meidinger}, N. and {Pfeffermann}, E. and {Reppin}, C. and {Aschenbach}, B. and {Bornemann}, W. and {Br{\"a}uninger}, H. and {Burkert}, W. and {Elender}, M. and {Freyberg}, M. and {Haberl}, F. and {Hartner}, G. and {Heuschmann}, F. and {Hippmann}, H. and {Kastelic}, E. and {Kemmer}, S. and {Kettenring}, G. and {Kink}, W. and {Krause}, N. and {M{\"u}ller}, S. and {Oppitz}, A. and {Pietsch}, W. and {Popp}, M. and {Predehl}, P. and {Read}, A. and {Stephan}, K.~H. and {St{\"o}tter}, D. and {Tr{\"u}mper}, J. and {Holl}, P. and {Kemmer}, J. and {Soltau}, H. and {St{\"o}tter}, R. and {Weber}, U. and {Weichert}, U. and {von Zanthier}, C. and {Carathanassis}, D. and {Lutz}, G. and {Richter}, R.~H. and {Solc}, P. and {B{\"o}ttcher}, H. and {Kuster}, M. and {Staubert}, R. and {Abbey}, A. and {Holland}, A. and {Turner}, M. and {Balasini}, M. and {Bignami}, G.~F. and {La Palombara}, N. and {Villa}, G. and {Buttler}, W. and {Gianini}, F. and {Lain{\'e}}, R. and {Lumb}, D. and {Dhez}, P.},
        title = "{The European Photon Imaging Camera on XMM-Newton: The pn-CCD camera}",
      journal = {\aap},
         year = 2001,
        month = jan,
       volume = {365},
        pages = {L18-L26},
          doi = {10.1051/0004-6361:20000066},
       adsurl = {https://ui.adsabs.harvard.edu/abs/2001A&A...365L..18S}
}

@ARTICLE{Jansen2001,
       author = {{Jansen}, F. and {Lumb}, D. and {Altieri}, B. and {Clavel}, J. and {Ehle}, M. and {Erd}, C. and {Gabriel}, C. and {Guainazzi}, M. and {Gondoin}, P. and {Much}, R. and {Munoz}, R. and {Santos}, M. and {Schartel}, N. and {Texier}, D. and {Vacanti}, G.},
        title = "{XMM-Newton observatory. I. The spacecraft and operations}",
      journal = {\aap},
         year = 2001,
        month = jan,
       volume = {365},
        pages = {L1-L6},
          doi = {10.1051/0004-6361:20000036},
       adsurl = {https://ui.adsabs.harvard.edu/abs/2001A&A...365L...1J}
}

@ARTICLE{Reynolds2024,
       author = {{Reynolds}, Mark and {Degenaar}, Natalie and {Wijnands}, Rudy and {Miller}, Jon and {Kennea}, Jamie},
        title = "{Swift GC monitoring program detection of low luminosity outburst a new source: Swift J174610-290018}",
      journal = {The Astronomer's Telegram},
         year = 2024,
        month = feb,
       volume = {16481},
        pages = {1},
       adsurl = {https://ui.adsabs.harvard.edu/abs/2024ATel16481....1R}
}

@ARTICLE{Grefenstette2025,
       author = {{Grefenstette}, B. and {Forster}, K. and {Zhang}, S. and {Mori}, K.},
        title = "{NuSTAR detects a source consistent with the location of the transient Swift J174610-290018}",
      journal = {The Astronomer's Telegram},
         year = 2025,
        month = apr,
       volume = {17132},
        pages = {1}
}

@ARTICLE{Pastor-Marazuela2020,
       author = {{Pastor-Marazuela}, I. and {Webb}, N.~A. and {Wojtowicz}, D.~T. and {van Leeuwen}, J.},
        title = "{The EXOD search for faint transients in XMM-Newton observations: Method and discovery of four extragalactic Type I X-ray bursters}",
      journal = {\aap},
         year = 2020,
        month = aug,
       volume = {640},
          eid = {A124},
        pages = {A124},
          doi = {10.1051/0004-6361/201936869},
archivePrefix = {arXiv},
       eprint = {2005.08673},
 primaryClass = {astro-ph.HE},
       adsurl = {https://ui.adsabs.harvard.edu/abs/2020A&A...640A.124P}
}

@ARTICLE{Jin2017,
       author = {{Jin}, Chichuan and {Ponti}, Gabriele and {Haberl}, Frank and {Smith}, Randall},
        title = "{Probing the interstellar dust towards the Galactic Centre: dust-scattering halo around AX J1745.6-2901}",
      journal = {\mnras},
         year = 2017,
        month = jul,
       volume = {468},
       number = {3},
        pages = {2532-2551},
          doi = {10.1093/mnras/stx653},
       adsurl = {https://ui.adsabs.harvard.edu/abs/2017MNRAS.468.2532J}
}

@ARTICLE{Jin2018,
       author = {{Jin}, Chichuan and {Ponti}, Gabriele and {Haberl}, Frank and {Smith}, Randall and {Valencic}, Lynne},
        title = "{Effects of interstellar dust scattering on the X-ray eclipses of the LMXB AX J1745.6-2901 in the Galactic Centre}",
      journal = {\mnras},
         year = 2018,
        month = jul,
       volume = {477},
       number = {3},
        pages = {3480-3506},
          doi = {10.1093/mnras/sty869},
       adsurl = {https://ui.adsabs.harvard.edu/abs/2018MNRAS.477.3480J}
}

@ARTICLE{Wilm2000,
       author = {{Wilms}, J. and {Allen}, A. and {McCray}, R.},
        title = "{On the Absorption of X-Rays in the Interstellar Medium}",
      journal = {\apj},
         year = 2000,
        month = oct,
       volume = {542},
       number = {2},
        pages = {914-924},
          doi = {10.1086/317016},
archivePrefix = {arXiv},
       eprint = {astro-ph/0008425},
 primaryClass = {astro-ph},
       adsurl = {https://ui.adsabs.harvard.edu/abs/2000ApJ...542..914W}
}

@INPROCEEDINGS{Arnaud1996,
       author = {{Arnaud}, K.~A.},
        title = "{XSPEC: The First Ten Years}",
    booktitle = {Astronomical Data Analysis Software and Systems V},
         year = 1996,
       editor = {{Jacoby}, George H. and {Barnes}, Jeannette},
       series = {Astronomical Society of the Pacific Conference Series},
       volume = {101},
        month = jan,
        pages = {17},
       adsurl = {https://ui.adsabs.harvard.edu/abs/1996ASPC..101...17A}
}

@ARTICLE{Buchner2014,
       author = {{Buchner}, J. and {Georgakakis}, A. and {Nandra}, K. and {Hsu}, L. and {Rangel}, C. and {Brightman}, M. and {Merloni}, A. and {Salvato}, M. and {Donley}, J. and {Kocevski}, D.},
        title = "{X-ray spectral modelling of the AGN obscuring region in the CDFS: Bayesian model selection and catalogue}",
      journal = {\aap},
         year = 2014,
        month = apr,
       volume = {564},
          eid = {A125},
        pages = {A125},
          doi = {10.1051/0004-6361/201322971},
archivePrefix = {arXiv},
       eprint = {1402.0004},
 primaryClass = {astro-ph.HE},
       adsurl = {https://ui.adsabs.harvard.edu/abs/2014A&A...564A.125B}
}

@ARTICLE{Degenaar2015,
       author = {{Degenaar}, N. and {Wijnands}, R. and {Miller}, J.~M. and {Reynolds}, M.~T. and {Kennea}, J. and {Gehrels}, N.},
        title = "{The Swift X-ray monitoring campaign of the center of the Milky Way}",
      journal = {Journal of High Energy Astrophysics},
         year = 2015,
        month = sep,
       volume = {7},
        pages = {137-147},
          doi = {10.1016/j.jheap.2015.03.005},
archivePrefix = {arXiv},
       eprint = {1503.07524},
 primaryClass = {astro-ph.HE},
       adsurl = {https://ui.adsabs.harvard.edu/abs/2015JHEAp...7..137D}
}

@ARTICLE{Wijnands2006,
       author = {{Wijnands}, R. and {in't Zand}, J.~J.~M. and {Rupen}, M. and {Maccarone}, T. and {Homan}, J. and {Cornelisse}, R. and {Fender}, R. and {Grindlay}, J. and {van der Klis}, M. and {Kuulkers}, E. and {Markwardt}, C.~B. and {Miller-Jones}, J.~C.~A. and {Wang}, Q.~D.},
        title = "{The XMM-Newton/Chandra monitoring campaign of the Galactic center region. Description of the program and preliminary results}",
      journal = {\aap},
         year = 2006,
        month = apr,
       volume = {449},
       number = {3},
        pages = {1117-1127},
          doi = {10.1051/0004-6361:20054129},
archivePrefix = {arXiv},
       eprint = {astro-ph/0508648},
 primaryClass = {astro-ph},
       adsurl = {https://ui.adsabs.harvard.edu/abs/2006A&A...449.1117W}
}

@ARTICLE{Zhu2018,
       author = {{Zhu}, Zhenlin and {Li}, Zhiyuan and {Morris}, Mark R.},
        title = "{An Ultradeep Chandra Catalog of X-Ray Point Sources in the Galactic Center Star Cluster}",
      journal = {\apjs},
         year = 2018,
        month = apr,
       volume = {235},
       number = {2},
          eid = {26},
        pages = {26},
          doi = {10.3847/1538-4365/aab14f},
archivePrefix = {arXiv},
       eprint = {1802.05073},
 primaryClass = {astro-ph.HE},
       adsurl = {https://ui.adsabs.harvard.edu/abs/2018ApJS..235...26Z}
}

@ARTICLE{Muno2005,
       author = {{Muno}, M.~P. and {Pfahl}, E. and {Baganoff}, F.~K. and {Brandt}, W.~N. and {Ghez}, A. and {Lu}, J. and {Morris}, M.~R.},
        title = "{An Overabundance of Transient X-Ray Binaries within 1 Parsec of the Galactic Center}",
      journal = {\apjl},
         year = 2005,
        month = apr,
       volume = {622},
       number = {2},
        pages = {L113-L116},
          doi = {10.1086/429721},
archivePrefix = {arXiv},
       eprint = {astro-ph/0412492},
 primaryClass = {astro-ph},
       adsurl = {https://ui.adsabs.harvard.edu/abs/2005ApJ...622L.113M}
}

@ARTICLE{Degenaar2012,
       author = {{Degenaar}, N. and {Wijnands}, R. and {Cackett}, E.~M. and {Homan}, J. and {in't Zand}, J.~J.~M. and {Kuulkers}, E. and {Maccarone}, T.~J. and {van der Klis}, M.},
        title = "{A four-year XMM-Newton/Chandra monitoring campaign of the Galactic centre: analysing the X-ray transients}",
      journal = {\aap},
         year = 2012,
        month = sep,
       volume = {545},
          eid = {A49},
        pages = {A49},
          doi = {10.1051/0004-6361/201219470},
archivePrefix = {arXiv},
       eprint = {1204.6043},
 primaryClass = {astro-ph.HE},
       adsurl = {https://ui.adsabs.harvard.edu/abs/2012A&A...545A..49D}
}

@ARTICLE{Hailey2018,
       author = {{Hailey}, Charles J. and {Mori}, Kaya and {Bauer}, Franz E. and {Berkowitz}, Michael E. and {Hong}, Jaesub and {Hord}, Benjamin J.},
        title = "{A density cusp of quiescent X-ray binaries in the central parsec of the Galaxy}",
      journal = {\nat},
         year = 2018,
        month = apr,
       volume = {556},
       number = {7699},
        pages = {70-73},
          doi = {10.1038/nature25029},
       adsurl = {https://ui.adsabs.harvard.edu/abs/2018Natur.556...70H}
}

@ARTICLE{Mori2021,
       author = {{Mori}, Kaya and {Hailey}, Charles J. and {Schutt}, Theo Y.~E. and {Mandel}, Shifra and {Heuer}, Keri and {Grindlay}, Jonathan E. and {Hong}, Jaesub and {Ponti}, Gabriele and {Tomsick}, John A.},
        title = "{The X-Ray Binary Population in the Galactic Center Revealed through Multi-decade Observations}",
      journal = {\apj},
         year = 2021,
        month = nov,
       volume = {921},
       number = {2},
          eid = {148},
        pages = {148},
          doi = {10.3847/1538-4357/ac1da5},
archivePrefix = {arXiv},
       eprint = {2108.07312},
 primaryClass = {astro-ph.HE},
       adsurl = {https://ui.adsabs.harvard.edu/abs/2021ApJ...921..148M}
}

@ARTICLE{Heinke2015,
       author = {{Heinke}, C.~O. and {Bahramian}, A. and {Degenaar}, N. and {Wijnands}, R.},
        title = "{The nature of very faint X-ray binaries: hints from light curves}",
      journal = {\mnras},
         year = 2015,
        month = mar,
       volume = {447},
       number = {4},
        pages = {3034-3043},
          doi = {10.1093/mnras/stu2652},
archivePrefix = {arXiv},
       eprint = {1412.4097},
 primaryClass = {astro-ph.HE},
       adsurl = {https://ui.adsabs.harvard.edu/abs/2015MNRAS.447.3034H}
}

@ARTICLE{Eijnden2018,
       author = {{van den Eijnden}, J. and {Degenaar}, N. and {Pinto}, C. and {Patruno}, A. and {Wette}, K. and {Messenger}, C. and {Hern{\'a}ndez Santisteban}, J.~V. and {Wijnands}, R. and {Miller}, J.~M. and {Altamirano}, D. and {Paerels}, F. and {Chakrabarty}, D. and {Fabian}, A.~C.},
        title = "{The very faint X-ray binary IGR J17062-6143: a truncated disc, no pulsations, and a possible outflow}",
      journal = {\mnras},
         year = 2018,
        month = apr,
       volume = {475},
       number = {2},
        pages = {2027-2044},
          doi = {10.1093/mnras/stx3224},
       adsurl = {https://ui.adsabs.harvard.edu/abs/2018MNRAS.475.2027V}
}

@ARTICLE{Degenaar2014,
       author = {{Degenaar}, N. and {Wijnands}, R. and {Reynolds}, M.~T. and {Miller}, J.~M. and {Altamirano}, D. and {Kennea}, J. and {Gehrels}, N. and {Haggard}, D. and {Ponti}, G.},
        title = "{The Peculiar Galactic Center Neutron Star X-Ray Binary XMM J174457-2850.3}",
      journal = {\apj},
         year = 2014,
        month = sep,
       volume = {792},
       number = {2},
          eid = {109},
        pages = {109},
          doi = {10.1088/0004-637X/792/2/109},
archivePrefix = {arXiv},
       eprint = {1406.4508},
 primaryClass = {astro-ph.HE},
       adsurl = {https://ui.adsabs.harvard.edu/abs/2014ApJ...792..109D}
}

@ARTICLE{King2006,
       author = {{King}, A.~R. and {Wijnands}, R.},
        title = "{The faintest accretors}",
      journal = {\mnras},
         year = 2006,
        month = feb,
       volume = {366},
       number = {1},
        pages = {L31-L34},
          doi = {10.1111/j.1745-3933.2005.00126.x},
archivePrefix = {arXiv},
       eprint = {astro-ph/0511486},
 primaryClass = {astro-ph},
       adsurl = {https://ui.adsabs.harvard.edu/abs/2006MNRAS.366L..31K}
}

@ARTICLE{Hameury2016,
       author = {{Hameury}, J. -M. and {Lasota}, J. -P.},
        title = "{Outbursts in ultracompact X-ray binaries}",
      journal = {\aap},
         year = 2016,
        month = oct,
       volume = {594},
          eid = {A87},
        pages = {A87},
          doi = {10.1051/0004-6361/201628434},
archivePrefix = {arXiv},
       eprint = {1607.06394},
 primaryClass = {astro-ph.HE},
       adsurl = {https://ui.adsabs.harvard.edu/abs/2016A&A...594A..87H}
}

@ARTICLE{Delsanto2007,
       author = {{Del Santo}, M. and {Sidoli}, L. and {Mereghetti}, S. and {Bazzano}, A. and {Tarana}, A. and {Ubertini}, P.},
        title = "{XMMU J174716.1-281048: a ``quasi-persistent'' very faint X-ray transient?}",
      journal = {\aap},
         year = 2007,
        month = jun,
       volume = {468},
       number = {1},
        pages = {L17-L20},
          doi = {10.1051/0004-6361:20077536},
archivePrefix = {arXiv},
       eprint = {0704.2134},
 primaryClass = {astro-ph},
       adsurl = {https://ui.adsabs.harvard.edu/abs/2007A&A...468L..17D}
}

@ARTICLE{Gravity2019,
       author = {{GRAVITY Collaboration} and {Abuter}, R. and {Amorim}, A. and {Baub{\"o}ck}, M. and {Berger}, J.~P. and {Bonnet}, H. and {Brandner}, W. and {Cl{\'e}net}, Y. and {Coud{\'e} Du Foresto}, V. and {de Zeeuw}, P.~T. and {Dexter}, J. and {Duvert}, G. and {Eckart}, A. and {Eisenhauer}, F. and {F{\"o}rster Schreiber}, N.~M. and {Garcia}, P. and {Gao}, F. and {Gendron}, E. and {Genzel}, R. and {Gerhard}, O. and {Gillessen}, S. and {Habibi}, M. and {Haubois}, X. and {Henning}, T. and {Hippler}, S. and {Horrobin}, M. and {Jim{\'e}nez-Rosales}, A. and {Jocou}, L. and {Kervella}, P. and {Lacour}, S. and {Lapeyr{\`e}re}, V. and {Le Bouquin}, J. -B. and {L{\'e}na}, P. and {Ott}, T. and {Paumard}, T. and {Perraut}, K. and {Perrin}, G. and {Pfuhl}, O. and {Rabien}, S. and {Rodriguez Coira}, G. and {Rousset}, G. and {Scheithauer}, S. and {Sternberg}, A. and {Straub}, O. and {Straubmeier}, C. and {Sturm}, E. and {Tacconi}, L.~J. and {Vincent}, F. and {von Fellenberg}, S. and {Waisberg}, I. and {Widmann}, F. and {Wieprecht}, E. and {Wiezorrek}, E. and {Woillez}, J. and {Yazici}, S.},
        title = "{A geometric distance measurement to the Galactic center black hole with 0.3\% uncertainty}",
      journal = {\aap},
         year = 2019,
        month = may,
       volume = {625},
          eid = {L10},
        pages = {L10},
          doi = {10.1051/0004-6361/201935656},
archivePrefix = {arXiv},
       eprint = {1904.05721},
 primaryClass = {astro-ph.GA},
       adsurl = {https://ui.adsabs.harvard.edu/abs/2019A&A...625L..10G}
}

@ARTICLE{Webb2020,
       author = {{Webb}, N.~A. and {Coriat}, M. and {Traulsen}, I. and {Ballet}, J. and {Motch}, C. and {Carrera}, F.~J. and {Koliopanos}, F. and {Authier}, J. and {de la Calle}, I. and {Ceballos}, M.~T. and {Colomo}, E. and {Chuard}, D. and {Freyberg}, M. and {Garcia}, T. and {Kolehmainen}, M. and {Lamer}, G. and {Lin}, D. and {Maggi}, P. and {Michel}, L. and {Page}, C.~G. and {Page}, M.~J. and {Perea-Calderon}, J.~V. and {Pineau}, F. -X. and {Rodriguez}, P. and {Rosen}, S.~R. and {Santos Lleo}, M. and {Saxton}, R.~D. and {Schwope}, A. and {Tom{\'a}s}, L. and {Watson}, M.~G. and {Zakardjian}, A.},
        title = "{The XMM-Newton serendipitous survey. IX. The fourth XMM-Newton serendipitous source catalogue}",
      journal = {\aap},
         year = 2020,
        month = sep,
       volume = {641},
          eid = {A136},
        pages = {A136},
          doi = {10.1051/0004-6361/201937353},
archivePrefix = {arXiv},
       eprint = {2007.02899},
 primaryClass = {astro-ph.HE},
       adsurl = {https://ui.adsabs.harvard.edu/abs/2020A&A...641A.136W}
}

@ARTICLE{Evans2024,
       author = {{Evans}, Ian N. and {Evans}, Janet D. and {Mart{\'\i}nez-Galarza}, J. Rafael and {Miller}, Joseph B. and {Primini}, Francis A. and {Azadi}, Mojegan and {Burke}, Douglas J. and {Civano}, Francesca M. and {D'Abrusco}, Raffaele and {Fabbiano}, Giuseppina and {Graessle}, Dale E. and {Grier}, John D. and {Houck}, John C. and {Lauer}, Jennifer and {McCollough}, Michael L. and {Nowak}, Michael A. and {Plummer}, David A. and {Rots}, Arnold H. and {Siemiginowska}, Aneta and {Tibbetts}, Michael S.},
        title = "{The Chandra Source Catalog Release 2 Series}",
      journal = {\apjs},
         year = 2024,
        month = oct,
       volume = {274},
       number = {2},
          eid = {22},
        pages = {22},
          doi = {10.3847/1538-4365/ad6319},
archivePrefix = {arXiv},
       eprint = {2407.10799},
 primaryClass = {astro-ph.HE},
       adsurl = {https://ui.adsabs.harvard.edu/abs/2024ApJS..274...22E}
}

@ARTICLE{Ponti2015,
       author = {{Ponti}, G. and {Morris}, M.~R. and {Terrier}, R. and {Haberl}, F. and {Sturm}, R. and {Clavel}, M. and {Soldi}, S. and {Goldwurm}, A. and {Predehl}, P. and {Nandra}, K. and {B{\'e}langer}, G. and {Warwick}, R.~S. and {Tatischeff}, V.},
        title = "{The XMM-Newton view of the central degrees of the Milky Way}",
      journal = {\mnras},
         year = 2015,
        month = oct,
       volume = {453},
       number = {1},
        pages = {172-213},
          doi = {10.1093/mnras/stv1331},
archivePrefix = {arXiv},
       eprint = {1508.04445},
 primaryClass = {astro-ph.HE},
       adsurl = {https://ui.adsabs.harvard.edu/abs/2015MNRAS.453..172P}
}

@ARTICLE{Wijnands2015,
       author = {{Wijnands}, R. and {Degenaar}, N. and {Armas Padilla}, M. and {Altamirano}, D. and {Cavecchi}, Y. and {Linares}, M. and {Bahramian}, A. and {Heinke}, C.~O.},
        title = "{Low-level accretion in neutron star X-ray binaries}",
      journal = {\mnras},
         year = 2015,
        month = dec,
       volume = {454},
       number = {2},
        pages = {1371-1386},
          doi = {10.1093/mnras/stv1974},
archivePrefix = {arXiv},
       eprint = {1409.6265},
 primaryClass = {astro-ph.HE},
       adsurl = {https://ui.adsabs.harvard.edu/abs/2015MNRAS.454.1371W}
}

@ARTICLE{Verner1996,
       author = {{Verner}, D.~A. and {Ferland}, G.~J. and {Korista}, K.~T. and {Yakovlev}, D.~G.},
        title = "{Atomic Data for Astrophysics. II. New Analytic Fits for Photoionization Cross Sections of Atoms and Ions}",
      journal = {\apj},
         year = 1996,
        month = jul,
       volume = {465},
        pages = {487},
          doi = {10.1086/177435},
archivePrefix = {arXiv},
       eprint = {astro-ph/9601009},
 primaryClass = {astro-ph},
       adsurl = {https://ui.adsabs.harvard.edu/abs/1996ApJ...465..487V}
}

@ARTICLE{Watson1981,
       author = {{Watson}, M.~G. and {Willingale}, R. and {Grindlay}, J.~E. and {Hertz}, P.},
        title = "{An X-ray study of the Galactic Center.}",
      journal = {\apj},
         year = 1981,
        month = nov,
       volume = {250},
        pages = {142-154},
          doi = {10.1086/159355},
       adsurl = {https://ui.adsabs.harvard.edu/abs/1981ApJ...250..142W}
}

@ARTICLE{Pavlinsky1994,
       author = {{Pavlinsky}, M.~N. and {Grebenev}, S.~A. and {Sunyaev}, R.~A.},
        title = "{X-Ray Images of the Galactic Center Obtained with Art-P/Granat: Discovery of New Sources, Variability of Persistent Sources, and Localization of X-Ray Bursters}",
      journal = {\apj},
         year = 1994,
        month = apr,
       volume = {425},
        pages = {110},
          doi = {10.1086/173967},
       adsurl = {https://ui.adsabs.harvard.edu/abs/1994ApJ...425..110P}
}

@ARTICLE{Sidoli1999,
       author = {{Sidoli}, L. and {Mereghetti}, S. and {Israel}, G.~L. and {Chiappetti}, L. and {Treves}, A. and {Orlandini}, M.},
        title = "{The Zoo of X-Ray Sources in the Galactic Center Region: Observations with BEPPOSAX}",
      journal = {\apj},
         year = 1999,
        month = nov,
       volume = {525},
       number = {1},
        pages = {215-227},
          doi = {10.1086/307887},
archivePrefix = {arXiv},
       eprint = {astro-ph/9904393},
 primaryClass = {astro-ph},
       adsurl = {https://ui.adsabs.harvard.edu/abs/1999ApJ...525..215S}
}

@ARTICLE{Sakano2002,
       author = {{Sakano}, Masaaki and {Koyama}, Katsuji and {Murakami}, Hiroshi and {Maeda}, Yoshitomo and {Yamauchi}, Shigeo},
        title = "{ASCA X-Ray Source Catalog in the Galactic Center Region}",
      journal = {\apjs},
         year = 2002,
        month = jan,
       volume = {138},
       number = {1},
        pages = {19-34},
          doi = {10.1086/324020},
archivePrefix = {arXiv},
       eprint = {astro-ph/0108376},
 primaryClass = {astro-ph},
       adsurl = {https://ui.adsabs.harvard.edu/abs/2002ApJS..138...19S}
}

@ARTICLE{Sidoli2001,
       author = {{Sidoli}, L. and {Belloni}, T. and {Mereghetti}, S.},
        title = "{A catalogue of soft X-ray sources in the galactic center region}",
      journal = {\aap},
         year = 2001,
        month = mar,
       volume = {368},
        pages = {835-844},
          doi = {10.1051/0004-6361:20010025},
archivePrefix = {arXiv},
       eprint = {astro-ph/0101156},
 primaryClass = {astro-ph},
       adsurl = {https://ui.adsabs.harvard.edu/abs/2001A&A...368..835S}
}

@ARTICLE{Kuulkers2007,
       author = {{Kuulkers}, E. and {Shaw}, S.~E. and {Paizis}, A. and {Chenevez}, J. and {Brandt}, S. and {Courvoisier}, T.~J. -L. and {Domingo}, A. and {Ebisawa}, K. and {Kretschmar}, P. and {Markwardt}, C.~B. and {Mowlavi}, N. and {Oosterbroek}, T. and {Orr}, A. and {R{\'\i}squez}, D. and {Sanchez-Fernandez}, C. and {Wijnands}, R.},
        title = "{The INTEGRAL Galactic bulge monitoring program: the first 1.5 years}",
      journal = {\aap},
         year = 2007,
        month = may,
       volume = {466},
       number = {2},
        pages = {595-618},
          doi = {10.1051/0004-6361:20066651},
archivePrefix = {arXiv},
       eprint = {astro-ph/0701244},
 primaryClass = {astro-ph},
       adsurl = {https://ui.adsabs.harvard.edu/abs/2007A&A...466..595K}
}

@ARTICLE{Muno2009,
       author = {{Muno}, M.~P. and {Bauer}, F.~E. and {Baganoff}, F.~K. and {Bandyopadhyay}, R.~M. and {Bower}, G.~C. and {Brandt}, W.~N. and {Broos}, P.~S. and {Cotera}, A. and {Eikenberry}, S.~S. and {Garmire}, G.~P. and {Hyman}, S.~D. and {Kassim}, N.~E. and {Lang}, C.~C. and {Lazio}, T.~J.~W. and {Law}, C. and {Mauerhan}, J.~C. and {Morris}, M.~R. and {Nagata}, T. and {Nishiyama}, S. and {Park}, S. and {Ram{\`\i}rez}, S.~V. and {Stolovy}, S.~R. and {Wijnands}, R. and {Wang}, Q.~D. and {Wang}, Z. and {Yusef-Zadeh}, F.},
        title = "{A Catalog of X-Ray Point Sources from Two Megaseconds of Chandra Observations of the Galactic Center}",
      journal = {\apjs},
         year = 2009,
        month = mar,
       volume = {181},
       number = {1},
        pages = {110-128},
          doi = {10.1088/0067-0049/181/1/110},
archivePrefix = {arXiv},
       eprint = {0809.1105},
 primaryClass = {astro-ph},
       adsurl = {https://ui.adsabs.harvard.edu/abs/2009ApJS..181..110M}
}

@ARTICLE{Muno2003,
       author = {{Muno}, M.~P. and {Baganoff}, F.~K. and {Bautz}, M.~W. and {Brandt}, W.~N. and {Broos}, P.~S. and {Feigelson}, E.~D. and {Garmire}, G.~P. and {Morris}, M.~R. and {Ricker}, G.~R. and {Townsley}, L.~K.},
        title = "{A Deep Chandra Catalog of X-Ray Point Sources toward the Galactic Center}",
      journal = {\apj},
         year = 2003,
        month = may,
       volume = {589},
       number = {1},
        pages = {225-241},
          doi = {10.1086/374639},
archivePrefix = {arXiv},
       eprint = {astro-ph/0301371},
 primaryClass = {astro-ph},
       adsurl = {https://ui.adsabs.harvard.edu/abs/2003ApJ...589..225M}
}

@INCOLLECTION{Bahramian2023,
       author = {{Bahramian}, Arash and {Degenaar}, Nathalie},
        title = "{Low-Mass X-ray Binaries}",
    booktitle = {Handbook of X-ray and Gamma-ray Astrophysics},
         year = 2023,
          eid = {120},
        pages = {120},
          doi = {10.1007/978-981-16-4544-0_94-1},
       adsurl = {https://ui.adsabs.harvard.edu/abs/2023hxga.book..120B}
}

@ARTICLE{Baganoff2003,
       author = {{Baganoff}, F.~K. and {Maeda}, Y. and {Morris}, M. and {Bautz}, M.~W. and {Brandt}, W.~N. and {Cui}, W. and {Doty}, J.~P. and {Feigelson}, E.~D. and {Garmire}, G.~P. and {Pravdo}, S.~H. and {Ricker}, G.~R. and {Townsley}, L.~K.},
        title = "{Chandra X-Ray Spectroscopic Imaging of Sagittarius A* and the Central Parsec of the Galaxy}",
      journal = {\apj},
         year = 2003,
        month = jul,
       volume = {591},
       number = {2},
        pages = {891-915},
          doi = {10.1086/375145},
archivePrefix = {arXiv},
       eprint = {astro-ph/0102151},
 primaryClass = {astro-ph},
       adsurl = {https://ui.adsabs.harvard.edu/abs/2003ApJ...591..891B}
}

@ARTICLE{Hofmann2018,
       author = {{Hofmann}, F. and {Ponti}, G. and {Haberl}, F. and {Clavel}, M.},
        title = "{New transient Galactic bulge intermediate polar candidate XMMU J175035.2-293557}",
      journal = {\aap},
         year = 2018,
        month = jul,
       volume = {615},
          eid = {L7},
        pages = {L7},
          doi = {10.1051/0004-6361/201832906},
archivePrefix = {arXiv},
       eprint = {1806.05526},
 primaryClass = {astro-ph.HE},
       adsurl = {https://ui.adsabs.harvard.edu/abs/2018A&A...615L...7H}
}

@ARTICLE{Mandel2025,
       author = {{Mandel}, Shifra and {Gerber}, Julian and {Mori}, Kaya and {Stringfield}, Ceaser and {Pe{\~n}aherrera}, Mabel and {Hailey}, Charles J. and {Du}, Alan and {Grindlay}, Jonathan and {Hong}, JaeSub and {Ponti}, Gabriele and {Tomsick}, John A. and {Berg}, Maureen van den},
        title = "{Hunting for High-mass X-Ray Binaries in the Galactic Center with NuSTAR}",
      journal = {\apj},
         year = 2025,
        month = jun,
       volume = {985},
       number = {2},
          eid = {202},
        pages = {202},
          doi = {10.3847/1538-4357/adc07e},
archivePrefix = {arXiv},
       eprint = {2503.21139},
 primaryClass = {astro-ph.HE},
       adsurl = {https://ui.adsabs.harvard.edu/abs/2025ApJ...985..202M}
}

@ARTICLE{Degenaar2017,
       author = {{Degenaar}, N. and {Pinto}, C. and {Miller}, J.~M. and {Wijnands}, R. and {Altamirano}, D. and {Paerels}, F. and {Fabian}, A.~C. and {Chakrabarty}, D.},
        title = "{An in-depth study of a neutron star accreting at low Eddington rate: on the possibility of a truncated disc and an outflow}",
      journal = {\mnras},
         year = 2017,
        month = jan,
       volume = {464},
       number = {1},
        pages = {398-409},
          doi = {10.1093/mnras/stw2355},
archivePrefix = {arXiv},
       eprint = {1609.04816},
 primaryClass = {astro-ph.HE},
       adsurl = {https://ui.adsabs.harvard.edu/abs/2017MNRAS.464..398D}
}

@ARTICLE{Fabian1989,
       author = {{Fabian}, A.~C. and {Rees}, M.~J. and {Stella}, L. and {White}, N.~E.},
        title = "{X-ray fluorescence from the inner disc in Cygnus X-1.}",
      journal = {\mnras},
         year = 1989,
        month = may,
       volume = {238},
        pages = {729-736},
          doi = {10.1093/mnras/238.3.729},
       adsurl = {https://ui.adsabs.harvard.edu/abs/1989MNRAS.238..729F}
}

@ARTICLE{Yoshimoto2025,
       author = {{Yoshimoto}, Anje and {Yamauchi}, Shigeo and {Nobukawa}, Masayoshi and {Uchiyama}, Hideki and {Nobukawa}, Kumiko K. and {Aoki}, Yuma and {Ishida}, Manabu and {Kanemaru}, Yoshiaki and {Shidatsu}, Megumi and {Hayashi}, Takayuki and {Maeda}, Yoshitomo and {Matsumoto}, Hironori and {Tsuboi}, Yohko and {Suzuki}, Hiromasa and {Nakajima}, Hiroshi and {Wang}, Q. Daniel and {Eguchi}, Satoshi and {Yoneyama}, Tomokage and {Dotani}, Tadayasu and {Behar}, Ehud and {Terada}, Yukikatsu and {Suzuki}, Nari and {Yoshimoto}, Marina},
        title = "{The unusual spectrum of the X-ray transient source XRISM J174610.8-290021 near the Galactic Center}",
      journal = {\pasj},
         year = 2025,
        month = jun,
       volume = {77},
          eid = {psaf063},
        pages = {psaf063},
          doi = {10.1093/pasj/psaf063},
archivePrefix = {arXiv},
       eprint = {2506.20088},
 primaryClass = {astro-ph.HE},
       adsurl = {https://ui.adsabs.harvard.edu/abs/2025PASJ...77S..96Y}
}

@ARTICLE{Kuulkers2003,
       author = {{Kuulkers}, E. and {den Hartog}, P.~R. and {in't Zand}, J.~J.~M. and {Verbunt}, F.~W.~M. and {Harris}, W.~E. and {Cocchi}, M.},
        title = "{Photospheric radius expansion X-ray bursts as standard candles}",
      journal = {\aap},
         year = 2003,
        month = feb,
       volume = {399},
        pages = {663-680},
          doi = {10.1051/0004-6361:20021781},
archivePrefix = {arXiv},
       eprint = {astro-ph/0212028},
 primaryClass = {astro-ph},
       adsurl = {https://ui.adsabs.harvard.edu/abs/2003A&A...399..663K}
}

@ARTICLE{Iaria2013,
       author = {{Iaria}, R. and {Di Salvo}, T. and {D'A{\`\i}}, A. and {Burderi}, L. and {Mineo}, T. and {Riggio}, A. and {Papitto}, A. and {Robba}, N.~R.},
        title = "{X-ray spectroscopy of the ADC source X1822-371 with Chandra and XMM-Newton}",
      journal = {\aap},
         year = 2013,
        month = jan,
       volume = {549},
          eid = {A33},
        pages = {A33},
          doi = {10.1051/0004-6361/201015305},
archivePrefix = {arXiv},
       eprint = {1210.0874},
 primaryClass = {astro-ph.HE},
       adsurl = {https://ui.adsabs.harvard.edu/abs/2013A&A...549A..33I}
}

@ARTICLE{Strohmayer2003,
      title={New Views of Thermonuclear Bursts}, 
      author={Tod Strohmayer and Lars Bildsten},
      year={2003},
      eprint={astro-ph/0301544},
      archivePrefix={arXiv},
      primaryClass={astro-ph},
      url={https://arxiv.org/abs/astro-ph/0301544}, 
}

@ARTICLE{Lewin1993,
       author = {{Lewin}, Walter H.~G. and {van Paradijs}, Jan and {Taam}, Ronald E.},
        title = "{X-Ray Bursts}",
      journal = {\ssr},
         year = 1993,
        month = sep,
       volume = {62},
       number = {3-4},
        pages = {223-389},
          doi = {10.1007/BF00196124},
       adsurl = {https://ui.adsabs.harvard.edu/abs/1993SSRv...62..223L}
}

@ARTICLE{White1982,
       author = {{White}, N.~E. and {Holt}, S.~S.},
        title = "{Accretion disk coronae.}",
      journal = {\apj},
         year = 1982,
        month = jun,
       volume = {257},
        pages = {318-337},
          doi = {10.1086/159991},
       adsurl = {https://ui.adsabs.harvard.edu/abs/1982ApJ...257..318W}
}

@ARTICLE{Dove1997,
       author = {{Dove}, James B. and {Wilms}, Jorn and {Begelman}, Mitchell C.},
        title = "{Self-consistent Thermal Accretion Disk Corona Models for Compact Objects. I. Properties of the Corona and the Spectrum of Escaping Radiation}",
      journal = {\apj},
         year = 1997,
        month = oct,
       volume = {487},
       number = {2},
        pages = {747-758},
          doi = {10.1086/304632},
archivePrefix = {arXiv},
       eprint = {astro-ph/9705108},
 primaryClass = {astro-ph},
       adsurl = {https://ui.adsabs.harvard.edu/abs/1997ApJ...487..747D}
}

@ARTICLE{Galloway2008,
       author = {{Galloway}, Duncan K. and {Muno}, Michael P. and {Hartman}, Jacob M. and {Psaltis}, Dimitrios and {Chakrabarty}, Deepto},
        title = "{Thermonuclear (Type I) X-Ray Bursts Observed by the Rossi X-Ray Timing Explorer}",
      journal = {\apjs},
         year = 2008,
        month = dec,
       volume = {179},
       number = {2},
        pages = {360-422},
          doi = {10.1086/592044},
archivePrefix = {arXiv},
       eprint = {astro-ph/0608259},
 primaryClass = {astro-ph},
       adsurl = {https://ui.adsabs.harvard.edu/abs/2008ApJS..179..360G}
}

@ARTICLE{Degenaar2018,
       author = {{Degenaar}, Nathalie and {Ballantyne}, David R. and {Belloni}, Tomaso and {Chakraborty}, Manoneeta and {Chen}, Yu-Peng and {Ji}, Long and {Kretschmar}, Peter and {Kuulkers}, Erik and {Li}, Jian and {Maccarone}, Thomas J. and {Malzac}, Julien and {Zhang}, Shu and {Zhang}, Shuang-Nan},
        title = "{Accretion Disks and Coronae in the X-Ray Flashlight}",
      journal = {\ssr},
         year = 2018,
        month = feb,
       volume = {214},
       number = {1},
          eid = {15},
        pages = {15},
          doi = {10.1007/s11214-017-0448-3},
archivePrefix = {arXiv},
       eprint = {1711.06272},
 primaryClass = {astro-ph.HE},
       adsurl = {https://ui.adsabs.harvard.edu/abs/2018SSRv..214...15D}
}

@ARTICLE{Degenaar2009,
       author = {{Degenaar}, N. and {Wijnands}, R.},
        title = "{The behavior of subluminous X-ray transients near the Galactic center as observed using the X-ray telescope aboard Swift}",
      journal = {\aap},
         year = 2009,
        month = feb,
       volume = {495},
       number = {2},
        pages = {547-559},
          doi = {10.1051/0004-6361:200810654},
archivePrefix = {arXiv},
       eprint = {0807.3458},
 primaryClass = {astro-ph},
       adsurl = {https://ui.adsabs.harvard.edu/abs/2009A&A...495..547D}
}

@ARTICLE{Nogueras2021,
       author = {{Nogueras-Lara}, F. and {Sch{\"o}del}, R. and {Neumayer}, N.},
        title = "{GALACTICNUCLEUS: A high-angular-resolution JHK$_{s}$ imaging survey of the Galactic centre. IV. Extinction maps and de-reddened photometry}",
      journal = {\aap},
         year = 2021,
        month = sep,
       volume = {653},
          eid = {A133},
        pages = {A133},
          doi = {10.1051/0004-6361/202140996},
archivePrefix = {arXiv},
       eprint = {2107.00021},
 primaryClass = {astro-ph.GA},
       adsurl = {https://ui.adsabs.harvard.edu/abs/2021A&A...653A.133N}
}

@ARTICLE{Xu2019,
       author = {{Xu}, Xiao-jie and {Yu}, Zhuo-li and {Li}, Xiang-dong},
        title = "{The Fe Line Flux Ratio as a Diagnostic of the Maximum Temperature and the White Dwarf Mass of Cataclysmic Variables}",
      journal = {\apj},
         year = 2019,
        month = jun,
       volume = {878},
       number = {1},
          eid = {53},
        pages = {53},
          doi = {10.3847/1538-4357/ab1fe1},
archivePrefix = {arXiv},
       eprint = {1905.03399},
 primaryClass = {astro-ph.SR},
       adsurl = {https://ui.adsabs.harvard.edu/abs/2019ApJ...878...53X}
}

@ARTICLE{Mukai2017,
       author = {{Mukai}, K.},
        title = "{X-Ray Emissions from Accreting White Dwarfs: A Review}",
      journal = {\pasp},
         year = 2017,
        month = jun,
       volume = {129},
       number = {976},
        pages = {062001},
          doi = {10.1088/1538-3873/aa6736},
archivePrefix = {arXiv},
       eprint = {1703.06171},
 primaryClass = {astro-ph.HE},
       adsurl = {https://ui.adsabs.harvard.edu/abs/2017PASP..129f2001M}
}

@ARTICLE{Lutovinov2013,
       author = {{Lutovinov}, A.~A. and {Revnivtsev}, M.~G. and {Tsygankov}, S.~S. and {Krivonos}, R.~A.},
        title = "{Population of persistent high-mass X-ray binaries in the Milky Way}",
      journal = {\mnras},
         year = 2013,
        month = may,
       volume = {431},
       number = {1},
        pages = {327-341},
          doi = {10.1093/mnras/stt168},
archivePrefix = {arXiv},
       eprint = {1302.0728},
 primaryClass = {astro-ph.HE},
       adsurl = {https://ui.adsabs.harvard.edu/abs/2013MNRAS.431..327L}
}

@ARTICLE{Scargle2013,
       author = {{Scargle}, Jeffrey D. and {Norris}, Jay P. and {Jackson}, Brad and {Chiang}, James},
        title = "{Studies in Astronomical Time Series Analysis. VI. Bayesian Block Representations}",
      journal = {\apj},
         year = 2013,
        month = feb,
       volume = {764},
       number = {2},
          eid = {167},
        pages = {167},
          doi = {10.1088/0004-637X/764/2/167},
archivePrefix = {arXiv},
       eprint = {1207.5578},
 primaryClass = {astro-ph.IM},
       adsurl = {https://ui.adsabs.harvard.edu/abs/2013ApJ...764..167S}
}

@ARTICLE{Stel2025,
       author = {{Stel}, Giovanni and {Ponti}, Gabriele and {Haardt}, Francesco and {Sormani}, Mattia},
        title = "{25 years of XMM-Newton observations of the Sgr A complex: 3D distribution and internal structure of the clouds}",
      journal = {\aap},
         year = 2025,
        month = mar,
       volume = {695},
          eid = {A52},
        pages = {A52},
          doi = {10.1051/0004-6361/202451359},
archivePrefix = {arXiv},
       eprint = {2501.09737},
 primaryClass = {astro-ph.GA},
       adsurl = {https://ui.adsabs.harvard.edu/abs/2025A&A...695A..52S}
}

@ARTICLE{Lomb1976,
       author = {{Lomb}, N.~R.},
        title = "{Least-Squares Frequency Analysis of Unequally Spaced Data}",
      journal = {\apss},
         year = 1976,
        month = feb,
       volume = {39},
       number = {2},
        pages = {447-462},
          doi = {10.1007/BF00648343},
       adsurl = {https://ui.adsabs.harvard.edu/abs/1976Ap&SS..39..447L}
}

@ARTICLE{Scargle1982,
       author = {{Scargle}, J.~D.},
        title = "{Studies in astronomical time series analysis. II. Statistical aspects of spectral analysis of unevenly spaced data.}",
      journal = {\apj},
         year = 1982,
        month = dec,
       volume = {263},
        pages = {835-853},
          doi = {10.1086/160554},
       adsurl = {https://ui.adsabs.harvard.edu/abs/1982ApJ...263..835S}
}

@ARTICLE{Mereghetti2015,
       author = {{Mereghetti}, Sandro and {Pons}, Jos{\'e} A. and {Melatos}, Andrew},
        title = "{Magnetars: Properties, Origin and Evolution}",
      journal = {\ssr},
         year = 2015,
        month = oct,
       volume = {191},
       number = {1-4},
        pages = {315-338},
          doi = {10.1007/s11214-015-0146-y},
archivePrefix = {arXiv},
       eprint = {1503.06313},
 primaryClass = {astro-ph.HE},
       adsurl = {https://ui.adsabs.harvard.edu/abs/2015SSRv..191..315M}
}

@ARTICLE{Coti_Zelati2018,
       author = {{Coti Zelati}, Francesco and {Rea}, Nanda and {Pons}, Jos{\'e} A. and {Campana}, Sergio and {Esposito}, Paolo},
        title = "{Systematic study of magnetar outbursts}",
      journal = {\mnras},
         year = 2018,
        month = feb,
       volume = {474},
       number = {1},
        pages = {961-1017},
          doi = {10.1093/mnras/stx2679},
archivePrefix = {arXiv},
       eprint = {1710.04671},
 primaryClass = {astro-ph.HE},
       adsurl = {https://ui.adsabs.harvard.edu/abs/2018MNRAS.474..961C}
}

@ARTICLE{Maeda1996,
       author = {{Maeda}, Yoshitomo and {Koyama}, Katsuji and {Sakano}, Masaaki and {Takeshima}, Toshiaki and {Yamauchi}, Shigeo},
        title = "{A New Eclipsing X-Ray Burster near the Galactic Center: A Quiescent State of the Old Transient A1742-289}",
      journal = {\pasj},
         year = 1996,
        month = jun,
       volume = {48},
        pages = {417-423},
          doi = {10.1093/pasj/48.3.417},
       adsurl = {https://ui.adsabs.harvard.edu/abs/1996PASJ...48..417M}
}

@ARTICLE{Degenaar2024,
       author = {{Degenaar}, N. and {Reynolds}, M.~T. and {Miller}, J.~M. and {Wijnands}, R. and {Kennea}, J.~A.},
        title = "{Swift/XRT captures a bright X-ray flare from the Galactic center}",
      journal = {The Astronomer's Telegram},
         year = 2024,
        month = jun,
       volume = {16642},
        pages = {1},
       adsurl = {https://ui.adsabs.harvard.edu/abs/2024ATel16642....1D}
}

@ARTICLE{Alizai2023,
       author = {{Alizai}, K. and {Chenevez}, J. and {Cumming}, A. and {Degenaar}, N. and {Falanga}, M. and {Galloway}, D.~K. and {in't Zand}, J.~J.~M. and {Jaisawal}, G.~K. and {Keek}, L. and {Kuulkers}, E. and {Lampe}, N. and {Schatz}, H. and {Serino}, M.},
        title = "{A catalogue of unusually long thermonuclear bursts on neutron stars}",
      journal = {\mnras},
         year = 2023,
        month = may,
       volume = {521},
       number = {3},
        pages = {3608-3624},
          doi = {10.1093/mnras/stad374},
archivePrefix = {arXiv},
       eprint = {2308.03499},
 primaryClass = {astro-ph.HE},
       adsurl = {https://ui.adsabs.harvard.edu/abs/2023MNRAS.521.3608A}
}

@ARTICLE{Degenaar2025,
       author = {{Degenaar}, N. and {Reynolds}, M.~T. and {Wijnands}, R. and {Miller}, J.~M. and {Kennea}, J.~A.},
        title = "{New outburst of the Galactic center transient Swift J174535.5-285921 seen with Swift/XRT}",
      journal = {The Astronomer's Telegram},
         year = 2025,
        month = may,
       volume = {17192},
        pages = {1},
       adsurl = {https://ui.adsabs.harvard.edu/abs/2025ATel17192....1D}
}

@ARTICLE{Hua2025,
       author = {{Hua}, Ziqian and {Li}, Zhiyuan},
        title = "{Is the Peculiar Galactic Center Transient Swift J174610.4-290018 A Recurrent Nova?}",
      journal = {arXiv e-prints},
         year = 2025,
        month = sep,
          eid = {arXiv:2509.26446},
        pages = {arXiv:2509.26446},
          doi = {10.48550/arXiv.2509.26446},
archivePrefix = {arXiv},
       eprint = {2509.26446},
 primaryClass = {astro-ph.HE},
       adsurl = {https://ui.adsabs.harvard.edu/abs/2025arXiv250926446H}
}

@ARTICLE{Eggleton1983,
       author = {{Eggleton}, P.~P.},
        title = "{Aproximations to the radii of Roche lobes.}",
      journal = {\apj},
         year = 1983,
        month = may,
       volume = {268},
        pages = {368-369},
          doi = {10.1086/160960},
       adsurl = {https://ui.adsabs.harvard.edu/abs/1983ApJ...268..368E}
}

\end{document}